\pdfoutput=1
\documentclass[a4paper,11pt]{article}

\usepackage{multirow}

\newcommand{\be}{\begin{equation}}

\newcommand{\ee}{\end{equation}}
\newcommand{\bea}{\begin{eqnarray}}
\newcommand{\eea}{\end{eqnarray}}
\newcommand{\bi}{\begin{itemize}}
\newcommand{\ei}{\end{itemize}}
\newcommand{\ben}{\begin{enumerate}}
\newcommand{\een}{\end{enumerate}}

\newcommand{\lc}{\left[}
\newcommand{\rc}{\right]}
\newcommand{\lp}{\left(}
\newcommand{\rp}{\right)}

\def\bsp#1\esp{\begin{split}#1\end{split}}

\usepackage{jheppub,slashed}

\preprint{LPSC-14-031, CERN-PH-TH/2014-010}
\title{Characterizing New Physics with Polarized Beams at High-Energy Hadron Colliders}

\author[a,b]{Benjamin Fuks,}
\author[c]{Josselin Proudom,}
\author[a]{Juan Rojo,}
\author[c]{Ingo Schienbein}

\affiliation[a]{Theory Division, Physics Department, CERN, CH-1211 Geneva 23,
  Switzerland}
\affiliation[b]{Institut Pluridisciplinaire Hubert Curien/D\'epartement
  Recherches Subatomiques, Universit\'e de Strasbourg/CNRS-IN2P3, 23 rue du
  Loess, F-67037 Strasbourg, France}
\affiliation[c]{Laboratoire de Physique Subatomique et de Cosmologie, Universit\'e Grenoble-Alpes, CNRS/IN2P3,
53 avenue des Martyrs, 38026 Grenoble, France}

\emailAdd{benjamin.fuks@cern.ch}
\emailAdd{josselin.proudom@lpsc.in2p3.fr}
\emailAdd{juan.rojo@cern.ch}
\emailAdd{ingo.schienbein@lpsc.in2p3.fr}

\abstract{The TeV energy region is currently being explored by both
the ATLAS and CMS experiments of the Large Hadron Collider and
phenomena beyond the Standard Model are extensively searched for.
Large fractions of the parameter space of many models have already
been excluded, and the ranges covered by the searches
will certainly be increased by the
upcoming energy and luminosity upgrades.
If new physics has to be discovered in the forthcoming years,
the ultimate goal of the high-energy physics program will consist of
fully characterizing the newly-discovered
degrees of freedom in terms of properties such as their masses, spins and couplings.
The scope of this paper is to show how the availability
of polarized beams at high-energy proton-proton colliders could yield
a unique discriminating power between different beyond the Standard Model
scenarios. We first discuss in a model-independent way how this discriminating power
arises from the differences between polarized and unpolarized parton distribution functions.
We then demonstrate how polarized beams allow one not only to disentangle different
production mechanisms giving the same final-state signature, but also to obtain
information on the parameters of the hypothetical new physics sector of the theory.
This is illustrated in the case of a particular class of scenarios leading to monotop production.
We consider three specific models that could produce a monotop
signature in unpolarized proton collisions, and show how they could be distinguished
by means of single- and double-spin asymmetries in polarized collisions.
Our results are presented for both the Large Hadron Collider operating
at a center-of-mass energy of 14 TeV and a recently
proposed Future Circular Collider assumed to collide protons
at a center-of-mass energy of 100 TeV.
}


\begin{document} 
\maketitle
\flushbottom

\section{Introduction}

After three years of data-taking, the ATLAS and CMS experiments have already
probed quite extensively the TeV scale.
With the upcoming proton-proton runs at 13~TeV and 14~TeV and
the proposed high-luminosity upgrade of the Large Hadron Collider (LHC),
searches for new phenomena,
particles and interactions promise to survey an even wider portion of
the parameter space of a huge variety of beyond the Standard Model (BSM) scenarios.
In most of the studies, the relevant experimental analyses are motivated
by theoretical arguments, implying some key new physics final-state signatures that
should be looked for. However, those signatures are neither typical of a given theory,
nor of a given benchmark scenario of a specific model.
One of the most famous examples illustrating this fact
arises from the Minimal Supersymmetric Standard Model
(MSSM)~\cite{Nilles:1983ge, Haber:1984rc} and Universal
Extra  Dimensions models~\cite{Appelquist:2000nn}, which
both predict the pair production
of Standard Model partners followed by their cascade decay into a final
state enriched in charged leptons and jets, and containing in addition a large
amount of missing transverse energy.
Consequently, beyond discovery, the task of disentangling
BSM theories (and even different scenarios within a specific
theory) that share a common final-state signature is
known to be far from trivial.

Additionally to the LHC, there is another high-energy hadron collider
which is providing an impressive wealth of results.
The RHIC collider at the Brookhaven National Laboratory
has successfully operated in its polarized
proton-proton mode at 200~GeV and 500~GeV, collecting
data with an integrated luminosity of more than 1~fb$^{-1}$.
Although these polarized collisions are mainly dedicated to spin physics,
pioneering BSM studies have shown the non-negligible impact of beam
polarization to get a handle on (some of) the model
parameters of specific theories~\cite{Craigie:1983as, Taxil:1996vf,Virey:1998ny,
Taxil:2001ft, Gehrmann:2004xu, Bozzi:2004qq,
Debove:2008nr}.
Another interesting possibility that has been put forward 
was the study of anomalous $WW\gamma$ and $WWZ$ couplings in a polarized
upgrade of the Tevatron collider~\cite{Wiest:1995iz}.
In addition to the existing RHIC polarized proton collider, most of the
aforementioned studies have also considered possible polarization
upgrades of both the Tevatron and the LHC. However, although those upgrades
have been already discussed in the past and are perfectly
feasible~\cite{Baiod:1995eu, DeRoeck:1999kx}, they are quite unlikely to be realized.

In contrast, first discussions on a Future Circular Collider (FCC)
with a center-of-mass energy of 100 TeV are now starting.
Therefore, this is the right time to begin to present the
physics cases motivating different operating options of
such a machine, including a possible polarized mode.
This paper lies in this context, and intends to show
how polarized proton beams colliding at 14~TeV and 100~TeV could provide
an interesting way of disentangling new physics models featuring
the same final-state signature.
The key ingredient that underlies this idea is the difference between polarized
and unpolarized parton distribution functions (PDFs)
for a given flavor, that leads to quite different
spin asymmetries in polarized collisions depending on the initial-state partonic
production channel. This thus allows one to distinguish different BSM
physics scenarios which, characterized by different production mechanism,
provide the same final-state signature in unpolarized collisions.

For the sake of illustration, we focus on the investigation of the recently
proposed monotop signature~\cite{Andrea:2011ws, Agram:2013wda}, which
corresponds to the production of a single top in association
with missing transverse energy and no other particle. Monotops naturally
appear in several extensions of the Standard Model, like for example in
supersymmetric theories with $R$-parity violation (RPV) where they are issued from the
decay of a singly-produced top
squark~\cite{Berger:1999zt,Berger:2000zk, Desai:2010sq, Fuks:2012im}.
Besides this RPV mode, monotops can also be produced in various dark matter
models~\cite{Davoudiasl:2010am,Davoudiasl:2011fj,Kamenik:2011nb,Alvarez:2013jqa} where the monotop
state originates either from
the decay of a vector resonance, or from tree-level flavor-changing
neutral interactions with a particle giving rise to missing energy.
In the following, we discuss how the measurements of single-spin and
double-spin asymmetries at a polarized LHC or at a polarized FCC would
allow one to get additional information on the nature of the initial partons
at the origin of the monotop signal and show how this could be used
in order to constrain the underlying new physics scenario.

The paper is organized as follows: in Section~\ref{sec:lumipol}, we perform a detailed study of parton densities and
parton luminosities in the framework of polarized proton-proton collisions at
center-of-mass energies of 14~TeV and 100~TeV, showing the strength of spin
asymmetries to discriminate among different initial states. Then, we assume the
observation of a monotop excess in unpolarized proton-proton collisions and
illustrate in Section~\ref{sec:monotops} how spin asymmetries possibly
allow one to get information on the new physics scenarios that have yielded
the signal. Our conclusions are presented in Section~\ref{sec:conclusions}.

\section{Spin asymmetries
at polarized hadron colliders}

\label{sec:lumipol}

As has been mentioned in the introduction, polarized beams at
high-energy hadron colliders would provide a unique opportunity to
characterize any new physics signal that might have been
previously observed in unpolarized collisions.
This appealing possibility relies on the fact that polarized and
unpolarized parton-parton luminosities show quite different behaviors for
a given flavor combination.
Therefore, single- and double-spin asymmetries in polarized hadron collisions
can provide information on the initial partonic state of any given process,
thus allowing one to
disentangle  different beyond the Standard Model production scenarios that lead to the
same final state signatures.

In the next section we will exploit these remarkable
properties to distinguish between
new physics scenarios for monotop production at the LHC,
characterized by different initial state production mechanisms and thus
by different single- and double-spin asymmetries in polarized collisions.
However, before discussing specific models, it is instructive to first
 evaluate a variety of single- and double-spin asymmetries at the
level of parton luminosities rather than at the full hadronic cross section
level.
This approximation is useful since in many cases of interest the polarized and unpolarized
matrix elements are similar, and thus the 
spin asymmetries computed from the partonic
luminosities only already carry the bulk of the relevant physics which is
accessible experimentally via the hadron-level asymmetries.

\begin{figure}
\begin{center}
\epsfig{width=0.49\textwidth,figure=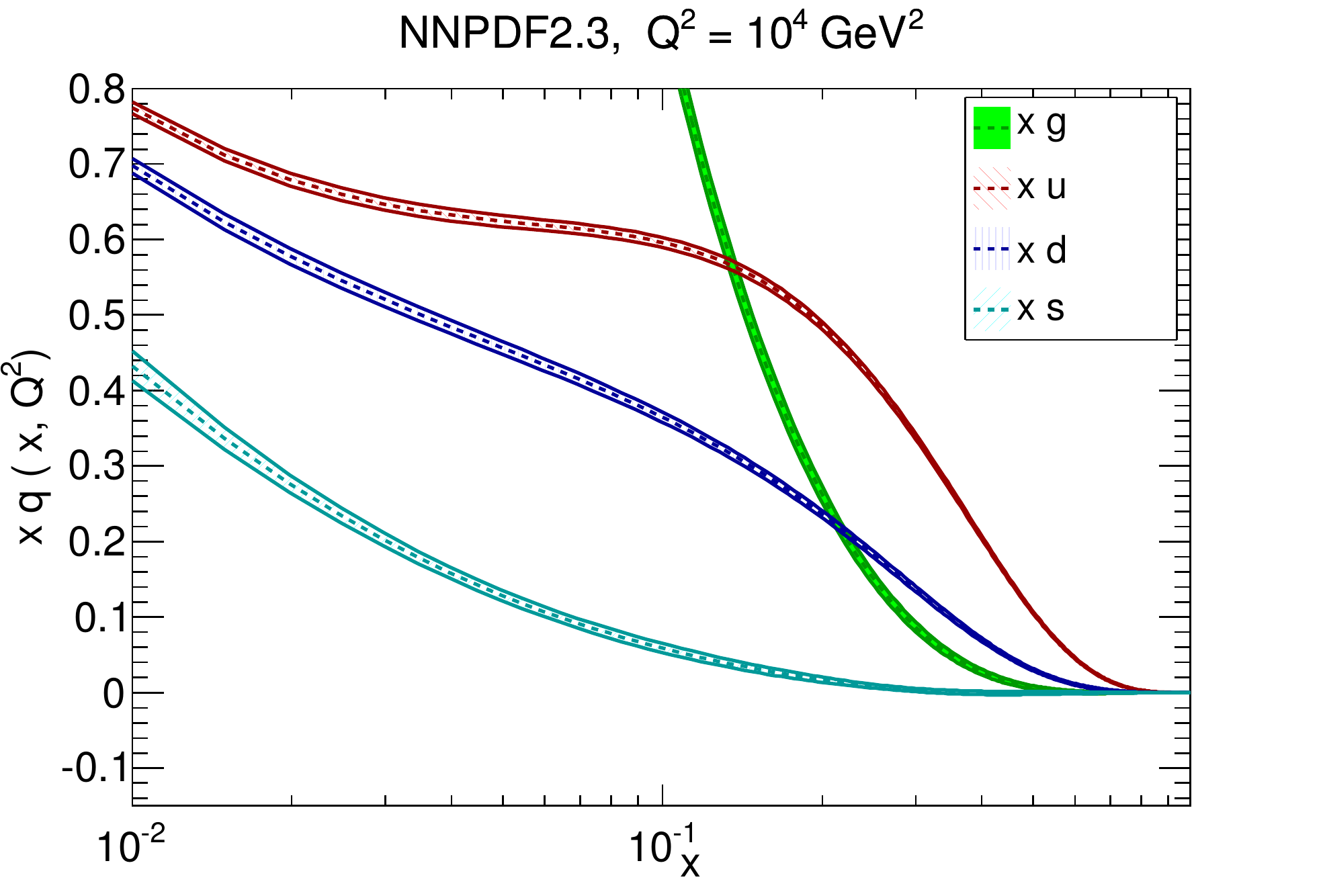}
\epsfig{width=0.49\textwidth,figure=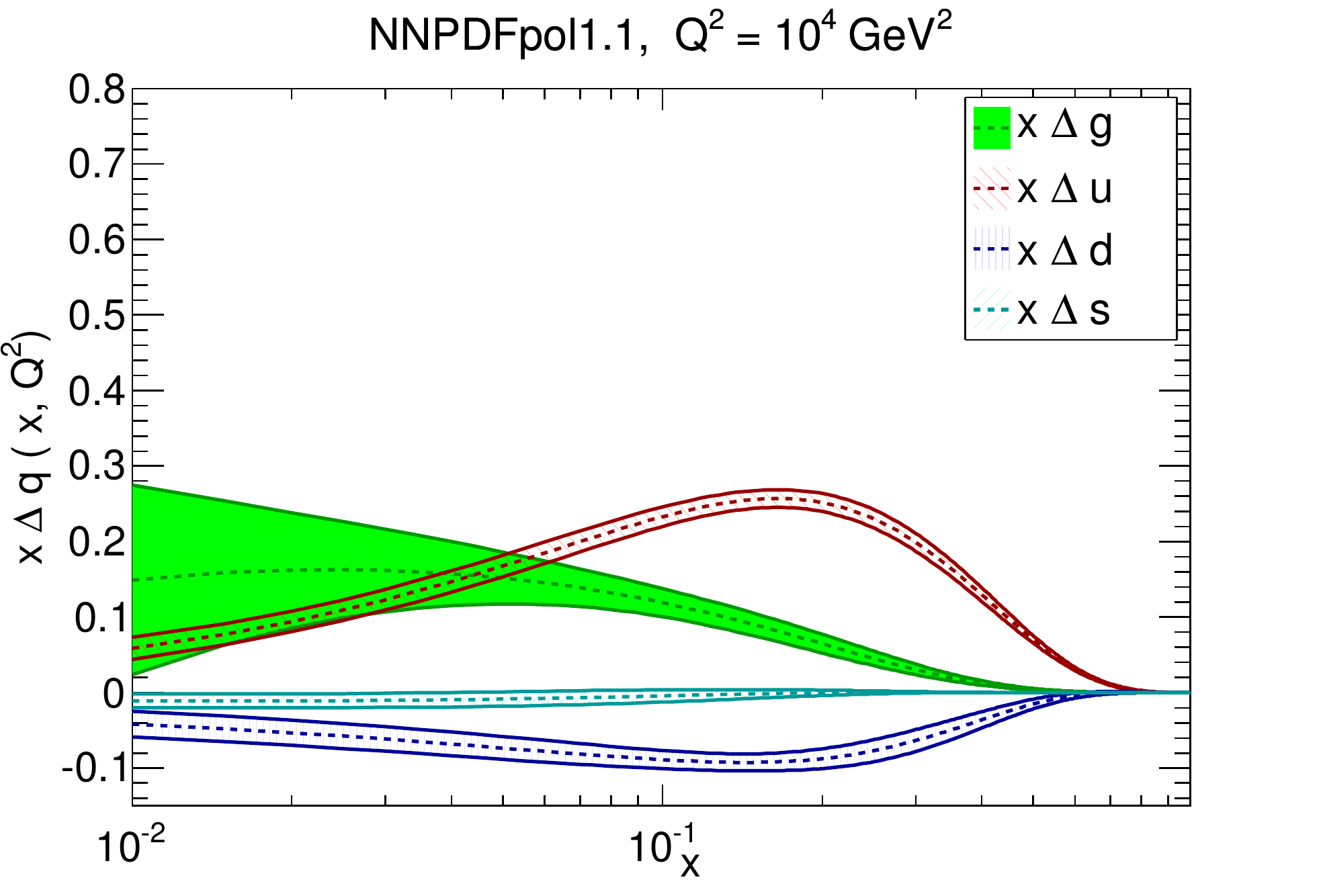}
\caption{\small Comparison between  unpolarized
(left) and polarized (right) up, down, strange
and gluon PDFs from the most
updated sets of the NNPDF family, NNPDF2.3 and NNPDFpol1.1 respectively.
PDFs have been evaluated at a typical high-energy hadron collider scale
of \mbox{$Q^2=10^4$~GeV$^2$}.
}
\label{fig:pdfcomp}
\end{center}
\end{figure}

First of all we compare polarized and unpolarized PDFs.
To fix the notation, we define unpolarized and polarized parton distributions
as usual,
\be
q_i(x,Q^2) \equiv q_i^{\uparrow}(x,Q^2) + q_i^{\downarrow}(x,Q^2) \ ,
\ee
\be
\Delta q_i(x,Q^2) \equiv q_i^{\uparrow}(x,Q^2) - q_i^{\downarrow}(x,Q^2) \ ,
\ee
in terms of the two different possible longitudinal polarization states
of partons within the nucleon,
\be
 q_i^{\uparrow}(x,Q^2) = q_i(x,Q^2)  + \Delta q_i(x,Q^2) \ ,
\ee
\be
 q_i^{\downarrow}(x,Q^2) = q_i(x,Q^2)  - \Delta q_i(x,Q^2) \ .
\ee
In Figure~\ref{fig:pdfcomp}, we present a comparison between the
different PDFs of the most updated unpolarized and polarized
sets from the NNPDF Collaboration\footnote{The polarized set of parton
densities NNPDFpol1.1 is available from the webpage
\texttt{https://nnpdf.hepforge.org/html/nnpdfpol10/nnpdfpol10sets.html}.},
NNPDF2.3~\cite{Ball:2012cx}
and NNPDFpol1.1~\cite{Ball:2013lla,Nocera:2014vla} respectively.
The various PDFs have been evaluated at a typical hadron collider scale of
$Q^2=10^4$ GeV$^2$ using the {\tt LHAPDF} interface~\cite{Whalley:2005nh}.

There are various interesting features to remark in Figure~\ref{fig:pdfcomp}.
The first one is that polarized PDFs are always smaller (in absolute value)
than their unpolarized counterparts, a consequence of the positivity
condition of polarized PDFs~\cite{Altarelli:1998gn}, which at Born
level reads,
\be
|\Delta q_i (x,Q^2) | \le q_i(x,Q^2) \ .
\ee
At the next-to-leading order, similar relations hold but only
for physical observables like polarized structure functions.
The second feature is that at small-$x$ the growth of
the polarized PDFs $x\Delta q_i(x,Q^2)$ is largely suppressed
with respect to that of the unpolarized ones $xq_i(x,Q^2)$~\cite{Ball:1995ye}.
As will be shown below, these two features have the important implication that
spin asymmetries will be sizable and thus experimentally
accessible only for final states with large invariant masses. This indeed probes
the polarized PDFs at medium and large values of $x$, two regions where
their magnitude is comparable to the one of the unpolarized parton densities.

In addition, and this is of particular importance for the problem at hand,
a specific flavor leads to
different qualitative behaviors for the polarized and unpolarized PDFs.
For instance, $\Delta u$ and $\Delta d$ have the opposite sign, while
$u$ and $d$ have both the same sign and the same shape.
This will translate into qualitatively different behaviors for the various
spin asymmetries depending on the underlying initial partonic state.

After comparing PDFs at the unpolarized and polarized level,
we move to the study of partonic luminosities~\cite{Campbell:2006wx}
and the corresponding single- and double-spin asymmetries.
We define the  partonic luminosity for the
scattering of two partons 
$i$ and $j$ in unpolarized
hadronic collisions, leading
to a final state of mass $m_X$, as
\begin{equation}
\label{eq:lumi1}
\mathcal{L}_{ij} = \frac{1}{S} \int_{\tau}^1 \frac{dx}{x}\frac{1}{1+\delta_{ij}}
\lc q_i\lp x,m_X\rp  q_j\lp \frac{\tau}{x},m_X\rp +
 q_i\lp \frac{\tau}{x},m_X\rp q_j\lp x,m_X\rp   \rc \, ,
\end{equation}
where the $\delta_{ij}$ factor removes the double counting in the case
of a same PDF combination $i=j$, and the collider center-of-mass
energy squared $S=E_{\rm cm}^2$ enters through the variable 
$\tau = \hat s/S$.
We can also define corresponding quantities involving polarized parton
distributions and thus relevant for polarized
collisions. The partonic luminosity relevant for
single-spin asymmetries is defined as
\be
\label{eq:lumi2}
\mathcal{L}^{L}_{ij}=\frac{1}{S} \int_{\tau}^1 \frac{dx}{x}\frac{1}{1+\delta_{ij}}
\lc q_i\lp x,m_X\rp  \Delta q_j\lp \frac{\tau}{x},m_X\rp +
 q_i\lp \frac{\tau}{x},m_X\rp \Delta q_j\lp x,m_X\rp   \rc \, ,
\ee
while for double-spin asymmetries, we use
\be
\label{eq:lumi3}
\mathcal{L}^{LL}_{ij}=\frac{1}{S} \int_{\tau}^1 \frac{dx}{x}\frac{1}{1+\delta_{ij}}
\lc \Delta q_i\lp x,m_X\rp  \Delta q_j\lp \frac{\tau}{x},m_X\rp +
\Delta q_i\lp \frac{\tau}{x},m_X\rp \Delta q_j\lp x,m_X\rp   \rc \ .
\ee
In our notation, the $L$ and $LL$ superscripts indicate that these luminosities
enter the description of
single- and double-spin asymmetries in polarized collisions,
respectively.

We now define unpolarized and polarized hadron-level cross sections by
\begin{eqnarray}
\sigma_0 & = & \frac{1}{4} 
\left[ 
\sigma^{\uparrow\uparrow}+\sigma^{\downarrow\downarrow}+ 
\sigma^{\uparrow\downarrow}+\sigma^{\downarrow\uparrow}
\right]\, ,
\label{eq:sig1}
\\
\sigma_L & = & \frac{1}{4}
\left[
\sigma^{\uparrow\uparrow}-\sigma^{\downarrow\downarrow}-
\sigma^{\uparrow\downarrow}+\sigma^{\downarrow\uparrow}
\right]\, ,
\label{eq:sig2}
\\
\sigma_{LL} &=&
\frac{1}{4} 
\left[ 
\sigma^{\uparrow\uparrow}+\sigma^{\downarrow\downarrow}-
\sigma^{\uparrow\downarrow}-\sigma^{\downarrow\uparrow}
\right]\, .
\label{eq:sig3}
\end{eqnarray}
Here $\sigma_0$ stands for the unpolarized cross sections and $\sigma_L$ and
$\sigma_{LL}$ for singly and doubly-polarized cross sections, respectively,
where an up-arrow denotes a helicity $h=+1$ and a down-arrow a helicity $h=-1$
of longitudinally polarized hadrons in the initial state. We recall that
in the case of singly-polarized cross sections, only one of the hadrons (the second
one here) is polarized.
It is useful to consider ratios of these cross sections, or spin asymmetries,
since theoretical and experimental uncertainties cancel to a good degree.
If a single beam is polarized, 
the experimentally relevant quantity is the single-spin asymmetry, defined as
\be
A_{L} = \frac{\sigma_L}{\sigma_0} \ ,
\label{eq:al1}
\ee
whereas if both beams are polarized, the relevant
quantity is the double-spin asymmetry
\be
A_{LL} = \frac{\sigma_{LL}}{\sigma_0} \ .
\label{eq:all1}
\ee

Eqs.~\eqref{eq:sig1}--\eqref{eq:sig3} define experimentally accessible
observables since they are expressed in terms of polarized hadrons.
In order to compare data with theoretical predictions,
in perturbative QCD the factorization theorem allows one to
write hadronic cross sections
as convolutions of parton distribution functions with parton level cross sections,
\begin{eqnarray}
\sigma_0 &=& q_i \otimes q_j \otimes \hat\sigma_{0,ij} = \mathcal{L}_{ij} \otimes [\hat s\ \hat\sigma_{0,ij}] \, ,
\label{eq:sigunpol}\\
\sigma_L &=& q_i \otimes \Delta q_j \otimes \hat\sigma_{L,ij} = \mathcal{L}^L_{ij} \otimes [\hat s\ \hat\sigma_{L,ij}] \, ,
\\
\sigma_{LL} &=& \Delta q_i \otimes \Delta q_j \otimes \hat\sigma_{LL,ij} = \mathcal{L}^{LL}_{ij} \otimes [\hat s\ \hat\sigma_{LL,ij}] \, .
\label{eq:sigLL}
\end{eqnarray}
The polarized partonic cross sections are here
defined in complete analogy to the polarized hadron-level expressions
of Eqs.~\eqref{eq:sig1}--\eqref{eq:sig3},
namely
\begin{eqnarray}
\hat\sigma_0 & = & \frac{1}{4} 
\left[ 
\hat\sigma^{\uparrow\uparrow}+\hat\sigma^{\downarrow\downarrow}+ 
\hat\sigma^{\uparrow\downarrow}+\hat\sigma^{\downarrow\uparrow}
\right]\, ,
\label{eq:sig1parton}
\\
\hat\sigma_L & = & \frac{1}{4}
\left[
\hat\sigma^{\uparrow\uparrow}-\hat\sigma^{\downarrow\downarrow}-
\hat\sigma^{\uparrow\downarrow}+\hat\sigma^{\downarrow\uparrow}
\right]\, ,
\label{eq:sig2parton}
\\
\hat\sigma_{LL} &=&
\frac{1}{4} 
\left[ 
\hat\sigma^{\uparrow\uparrow}+\hat\sigma^{\downarrow\downarrow}-
\hat\sigma^{\uparrow\downarrow}-\hat\sigma^{\downarrow\uparrow}
\right]\, ,
\label{eq:sig3parton}
\end{eqnarray}
where now the helicities are those of the incoming quarks and gluons
in the partonic collision.
Furthermore, a sum over all relevant partonic subprocesses is implied (\textit{i.e.}, over $i,j$)
and we refer to Eqs.~\eqref{eq:lumi1}--\eqref{eq:lumi3} for the definition of the
partonic luminosities.

For many cases of physical interest, the expressions in Eqs.\ \eqref{eq:sigunpol}--\eqref{eq:sigLL}
and consequently the asymmetries in Eqs.\ \eqref{eq:al1} and \eqref{eq:all1}
can be further simplified.
Firstly, the dimensionless cross sections $\hat s \hat \sigma_{ij}$ are often either constant far above the production threshold
(see, \textit{e.g.}, Figure~70 in Ref.~\cite{Campbell:2006wx})
or, in the case of a narrow $s$-channel resonance, they are peaked
at threshold, that is, $\hat s \simeq m_X^2$.
In the latter case, we end up having simple expressions of the hadron-level asymmetries
in terms of (ratios of weighted sums of) parton luminosities.
Secondly, the absolute values of the polarized and unpolarized parton-level matrix
elements are often the same or very similar, leading to further simplifications.
In cases where there is a single dominant particular sub-channel,
the hadronic asymmetries are just simple ratios of parton luminosities,
as can be deduced from the single-spin and
double-spin asymmetries of
Eqs.~\eqref{eq:sigunpol}--\eqref{eq:sigLL},
\be
\label{eq:al}
A_{L}^{ij} = \frac{\mathcal{L}^{L}_{ij}}{\mathcal{L}_{ij}}
\qquad\text{and}\qquad
A_{LL}^{ij} = \frac{\mathcal{L}^{LL}_{ij}}{\mathcal{L}_{ij}} \ .
\ee

In the rest of this section, we focus on
results for single and double-spin asymmetries computed
from Eq.~(\ref{eq:al}) for different initial state
partonic sub-channels.
We have calculated these asymmetries for the LHC collider operating
at a center-of-mass energy of 14~TeV (LHC 14 TeV), assuming a possible
future polarized upgrade, as well as for the polarized mode
of an hypothetical Future Circular Collider with a center-of-mass energy
of 100~TeV (FCC 100 TeV).
As polarized PDFs we use the NNPDFpol1.1~\cite{Ball:2013lla,Nocera:2014vla}
and DSSV08~\cite{deFlorian:2009vb} sets, together with the
corresponding unpolarized counterparts, NNPDF2.3~\cite{Ball:2012cx}
and MRST01~\cite{Martin:2001es}.
Comparing the predictions of NNPDFpol1.1 with those
of DSSV08 is useful in order to verify which features of the
spin asymmetries are generic irrespective of the specific details
of the particular polarized PDF set used.

It is clear from the definition of Eqs.~\eqref{eq:lumi1}--\eqref{eq:lumi3}
that to first approximation, luminosities are invariant
if the center-of-mass energy is modified, $\sqrt{S'}=k\sqrt{S}$, provided
that the final state mass is also modified in the same way,
$m_X'=km_X$, since in this case the variable $\tau$ is invariant.
However, logarithmic corrections to the DGLAP evolution of the PDFs modify this picture,
though they should not change any qualitative conclusion.
This property will be explicitly verified below when comparing the spin
asymmetries at \mbox{LHC 14 TeV} and at\mbox{ FCC 100 TeV}.

\begin{figure}
\begin{center}
\epsfig{width=0.32\textwidth,figure=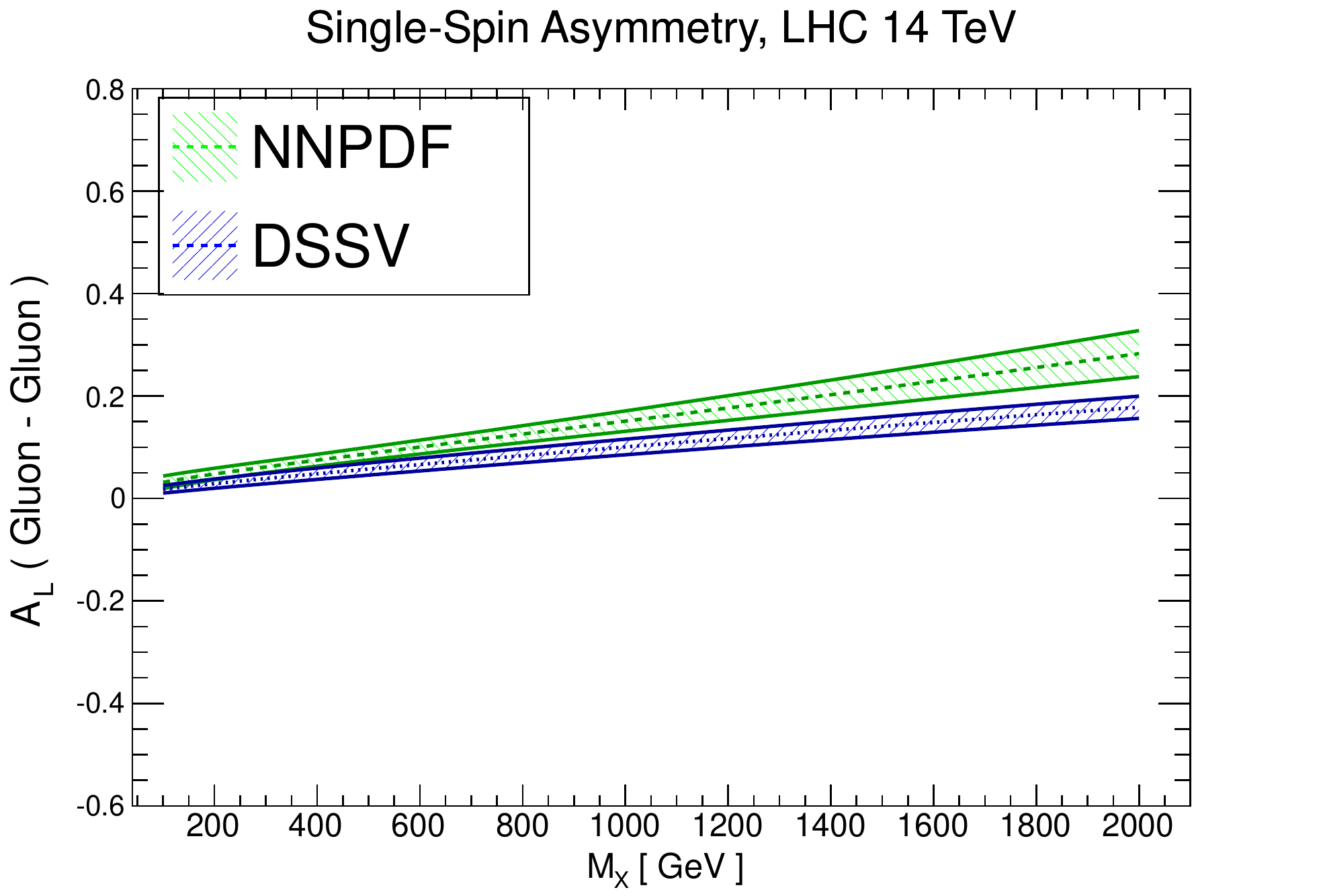}
\epsfig{width=0.32\textwidth,figure=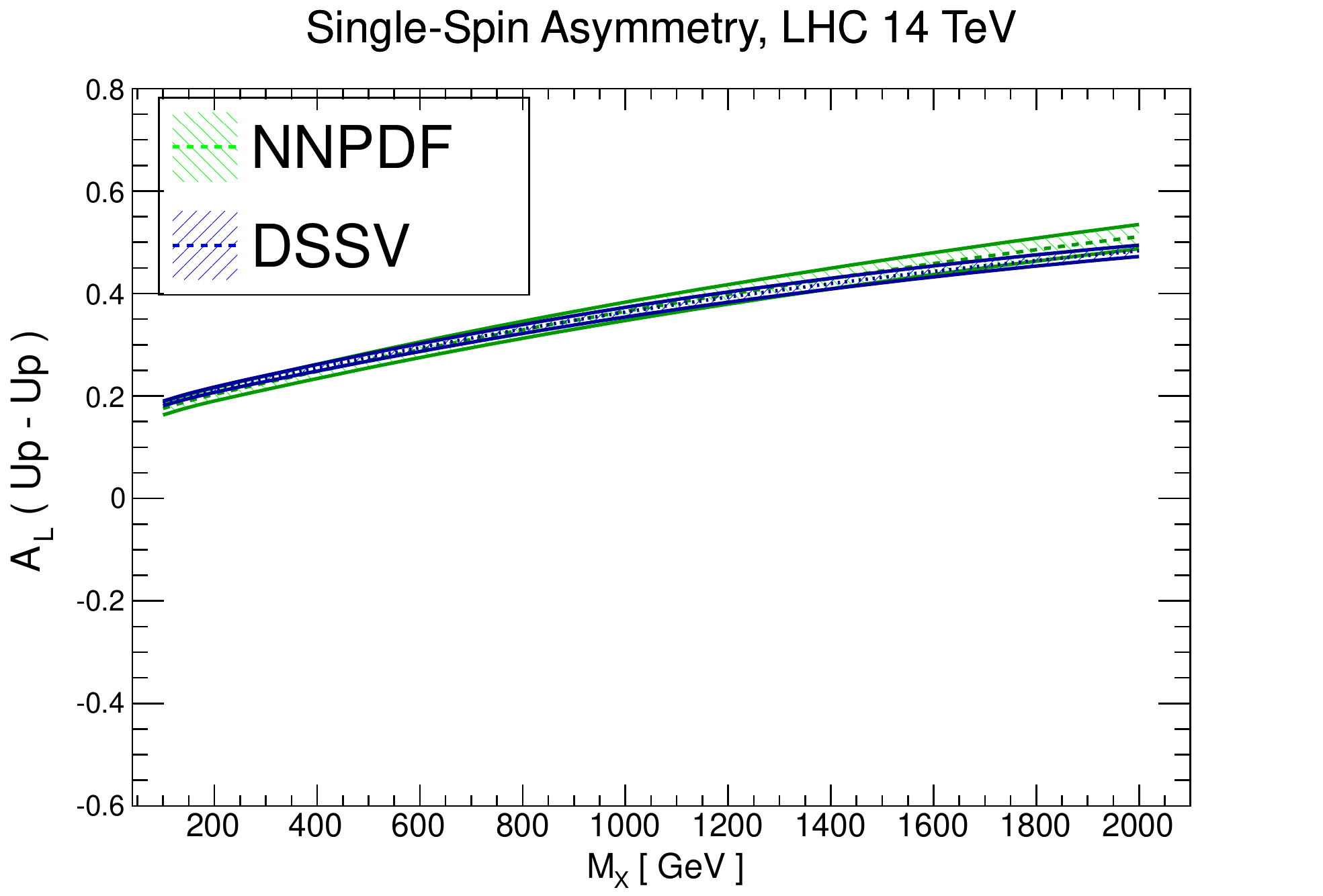}
\epsfig{width=0.32\textwidth,figure=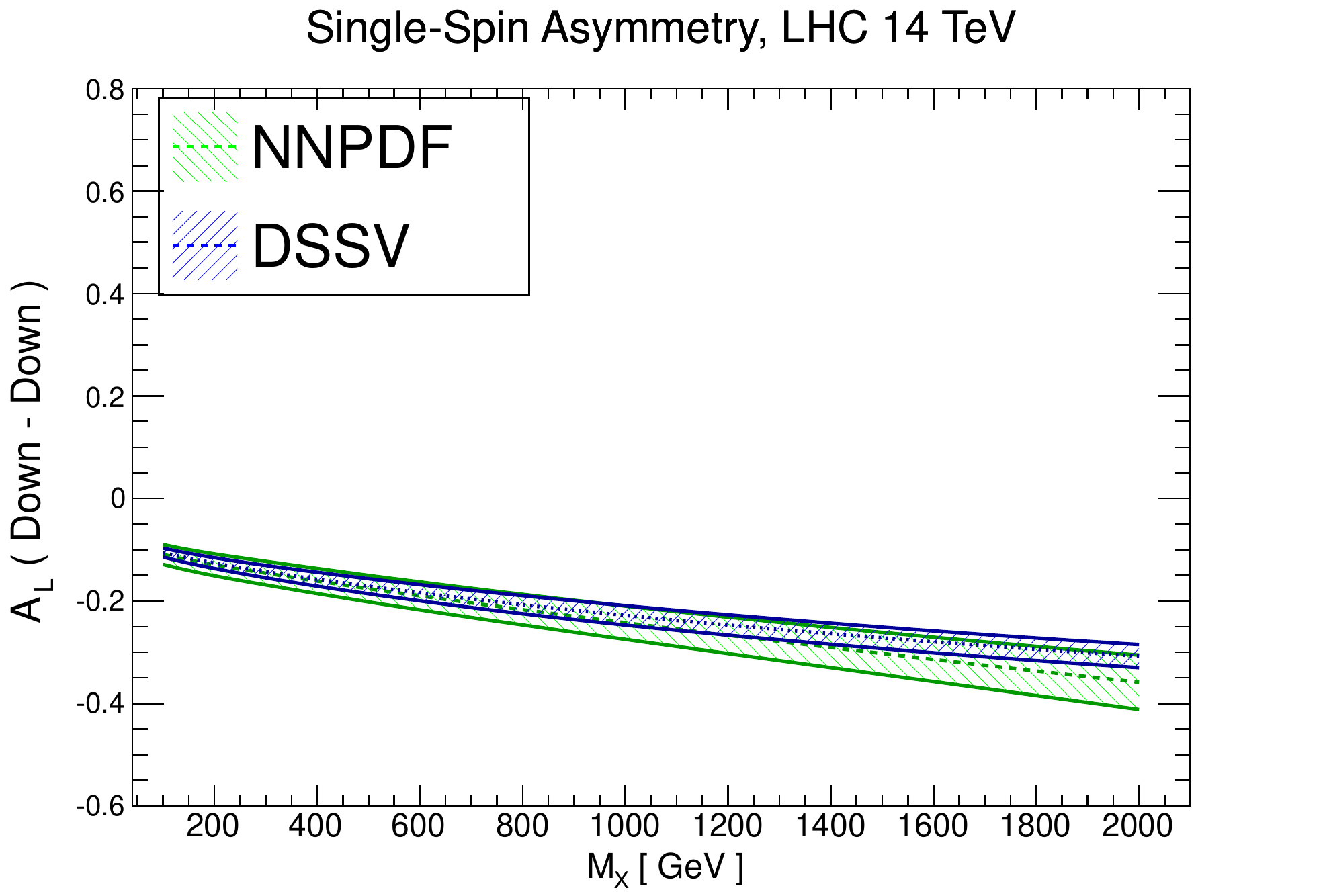} \\ \vspace{.4cm}
\epsfig{width=0.32\textwidth,figure=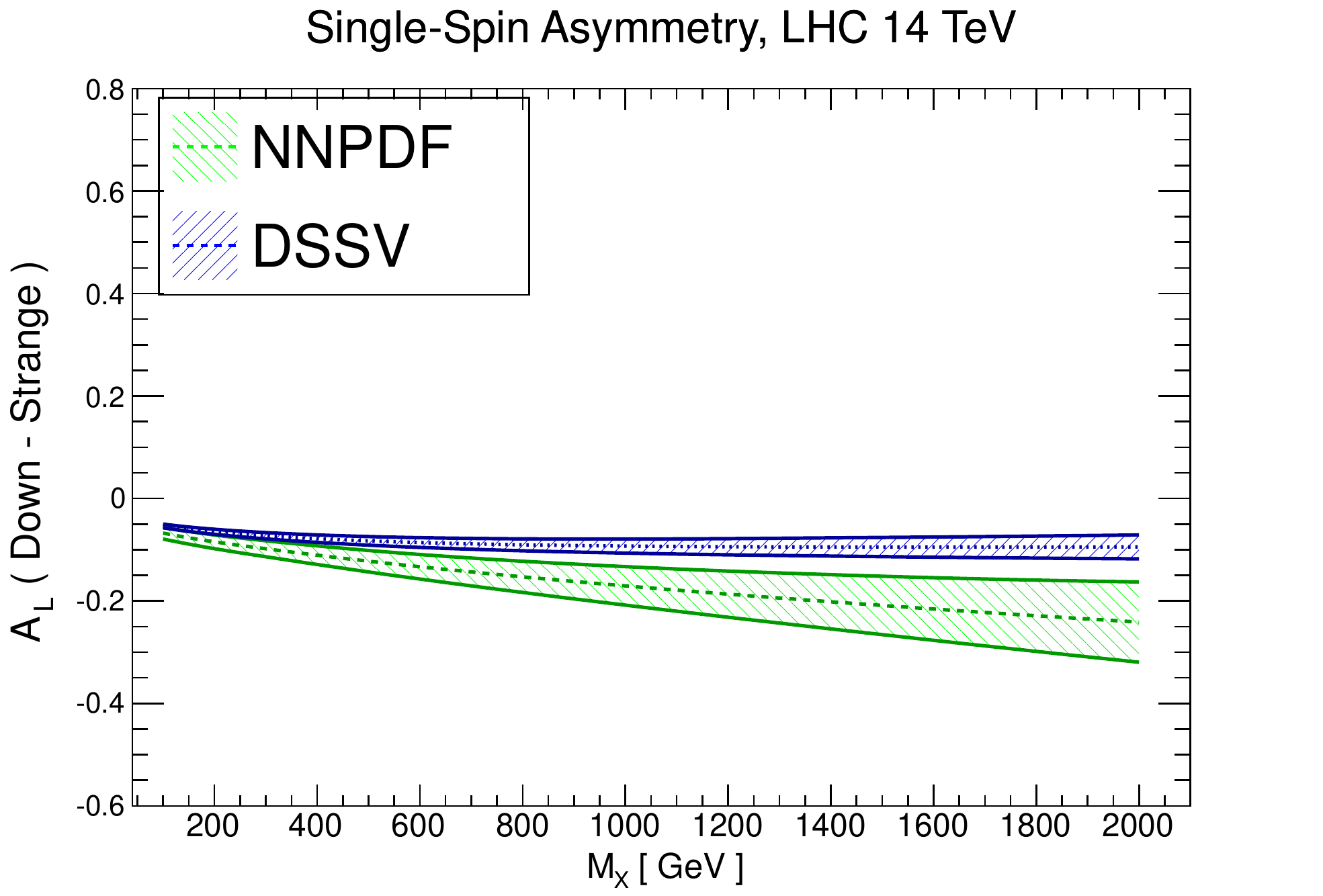}
\epsfig{width=0.32\textwidth,figure=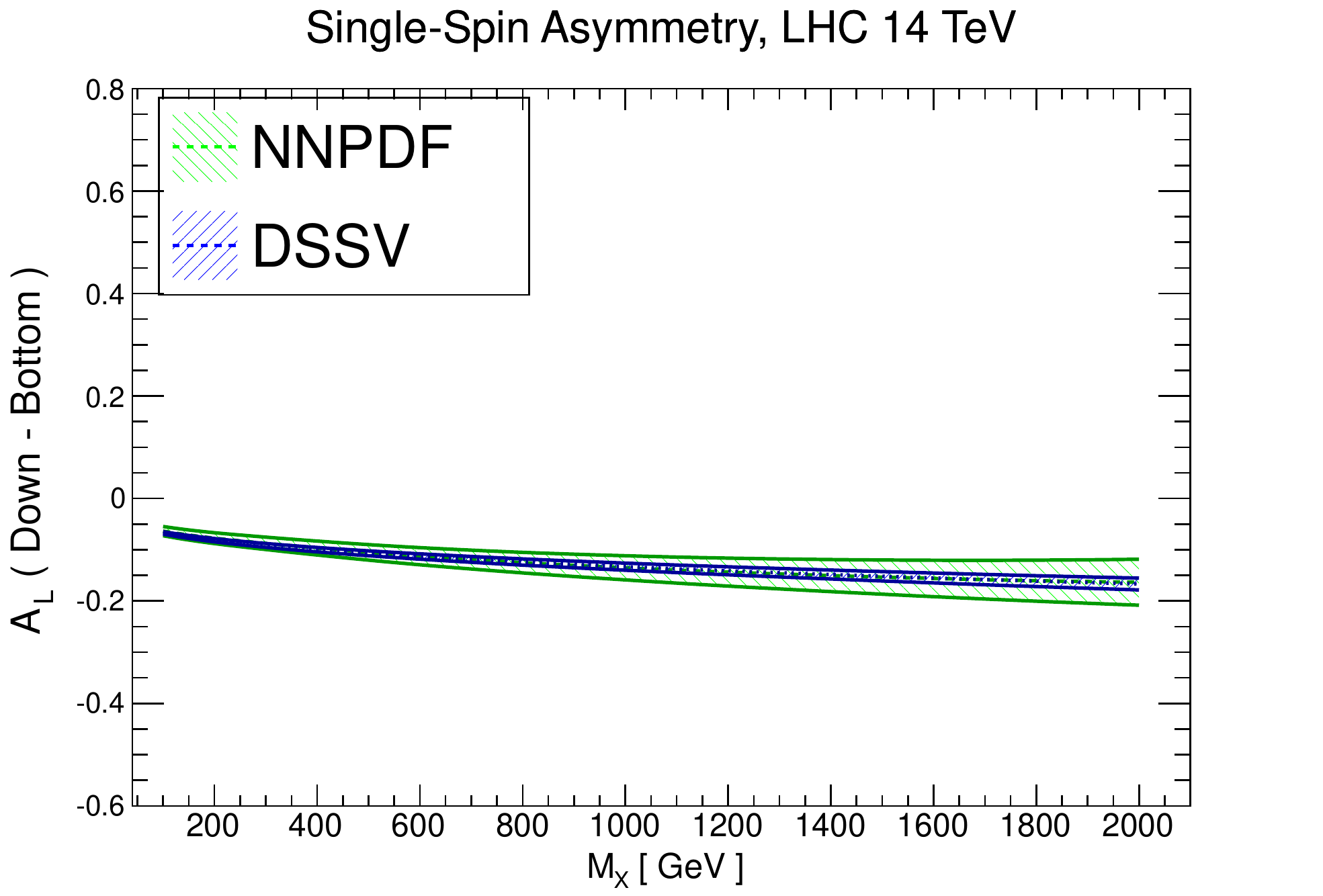}
\epsfig{width=0.32\textwidth,figure=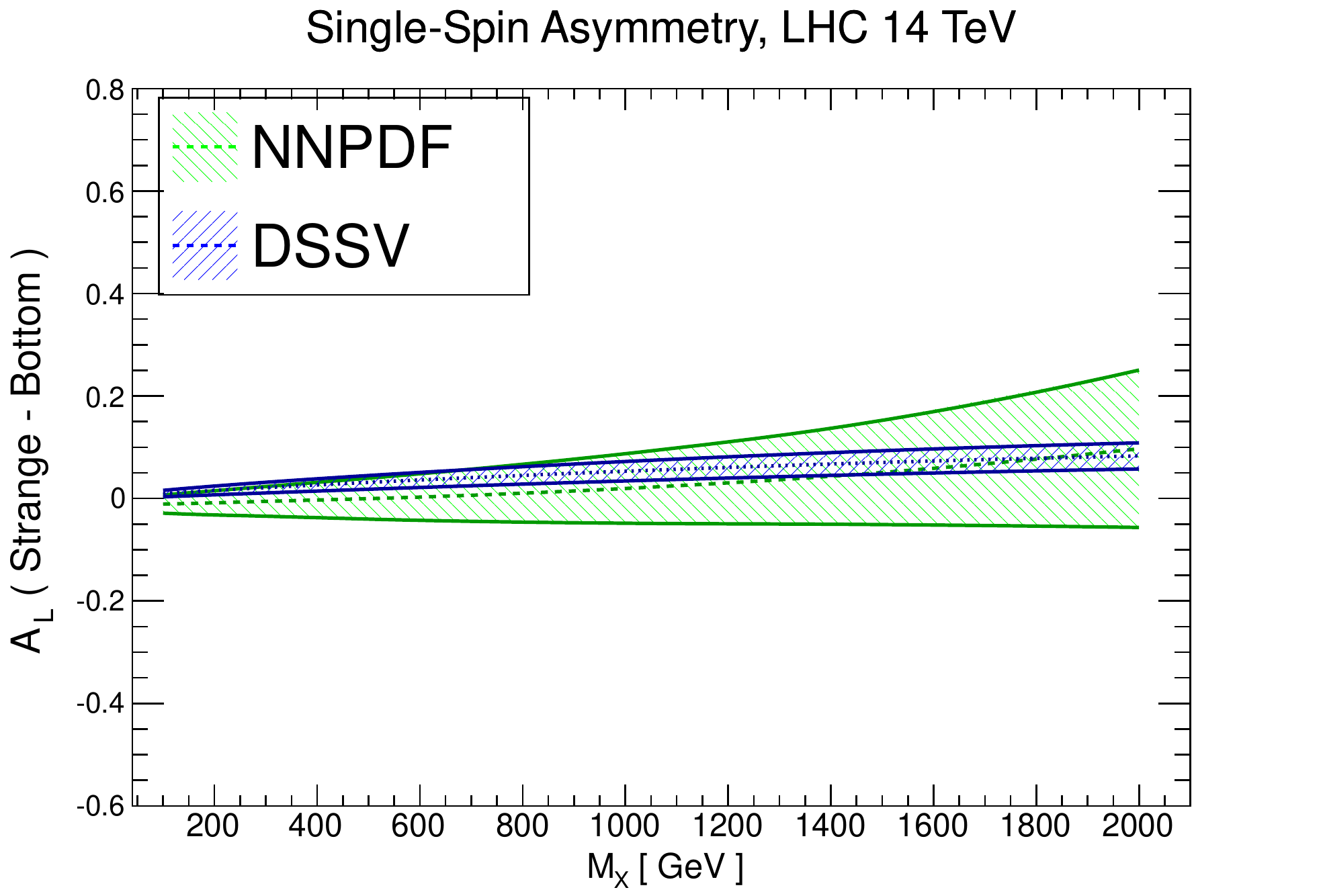}
\caption{\small The single-spin asymmetry $A_{L}$ at the parton luminosity level,
that we compute from Eq.~(\ref{eq:al}), at LHC 14 TeV, and for various initial-state
partonic combinations. We compare results obtained using NNPDFpol1.1/NNPDF2.3
with those obtained using DSSV/MRST and present them as function of the invariant mass
of the final state $m_X$.
The bands correspond to the polarized PDF uncertainties.
}
\label{fig:AL_LHC14}
\end{center}
\end{figure}

First of all, we compare the single-spin asymmetries at LHC 14 TeV for
the production of a final state with invariant mass $m_X$ assuming different
partonic initial states.
We compare the consistency of the asymmetries obtained with NNPDF with those
obtained with DSSV/MRST.
In all cases, the uncertainty band on the asymmetries corresponds to that of the
polarized PDFs, since in this respect the unpolarized PDF uncertainties can be 
neglected.
We show the asymmetries for $gg$, $uu$ and $dd$ initial states in the upper row
of Figure~\ref{fig:AL_LHC14}
and for the $ds$, $db$ and $sb$ initial states in the bottom row of the figure.
The DSSV08 densities consist of a PDF set obtained in the fixed-flavor-number scheme, and therefore
the polarized bottom PDF $\Delta b=\Delta\bar{b}=0$.
While differences between fixed-flavor-number and variable-flavor-number schemes
lead to substantial differences for unpolarized PDFs~\cite{Ball:2013gsa}, this is considered
 less important for polarized PDFs in the region with available experimental
data where the contribution from heavy quarks is small.
However, this is no longer true when evolving upwards in $Q^2$ to the
region relevant for collider physics, where heavy quark PDFs are not
negligible even in the polarized case.

In general there is a reasonable
qualitative agreement between the results from NNPDFpol1.1 and those of DSSV, with some
quantitative differences, for instance in asymmetries that involve the
polarized strange PDF.
This is expected since NNPDFpol1.1 and DSSV08 generally agree well for all
PDFs but for $\Delta s(x,Q^2)$, where even the sign is opposite~\cite{Ball:2013gsa}.
Larger PDF uncertainties are obtained using NNPDFpol1.1, partially due to the more
flexible functional form 
of the input PDFs as compared to DSSV08.
Results for the single
spin asymmetries for a 100 TeV FCC are qualitatively similar once the value of
the final state mass is properly rescaled as discussed above, so that results are not shown
explicitly.

\begin{figure}
\begin{center}
\epsfig{width=0.49\textwidth,figure=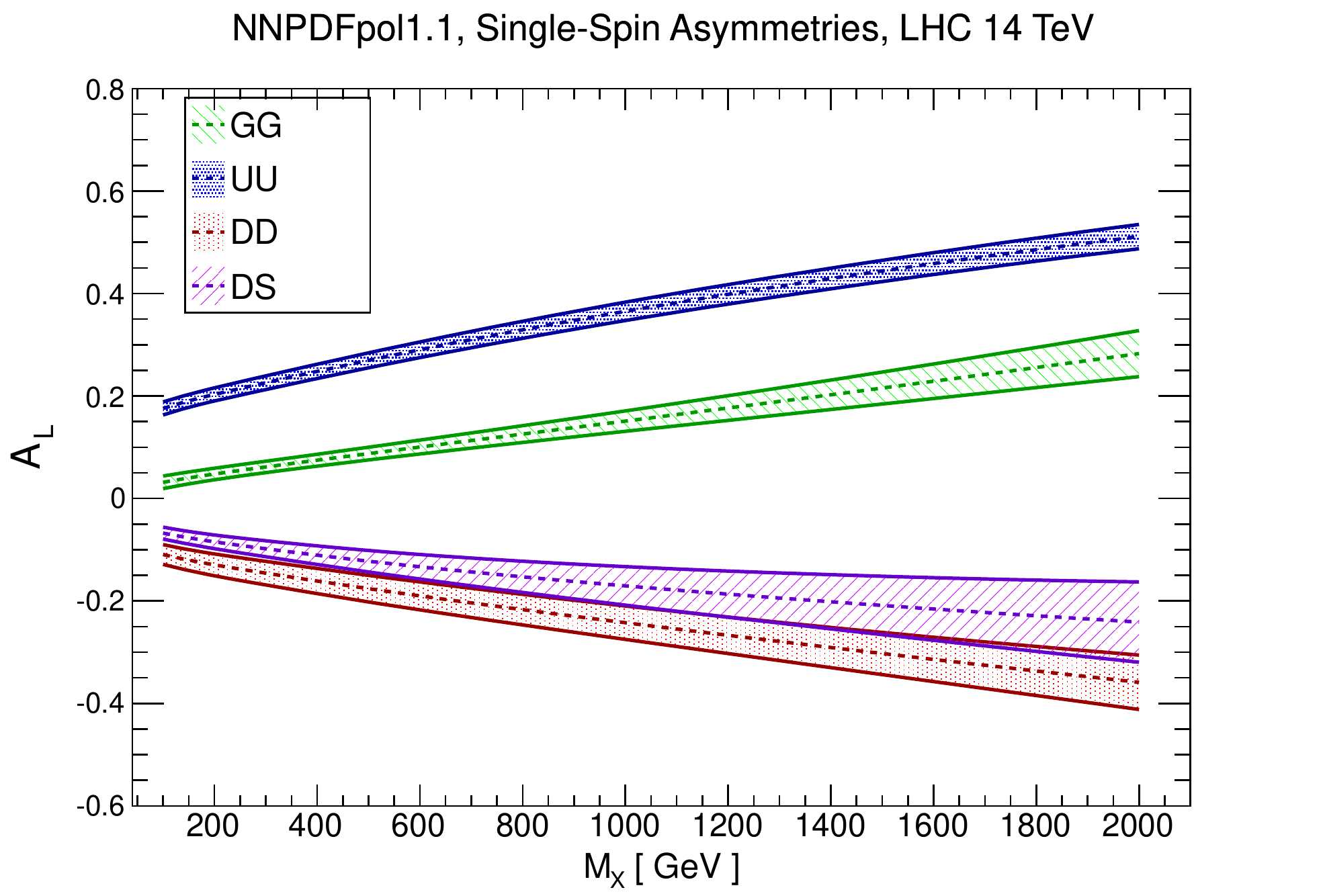}
\epsfig{width=0.49\textwidth,figure=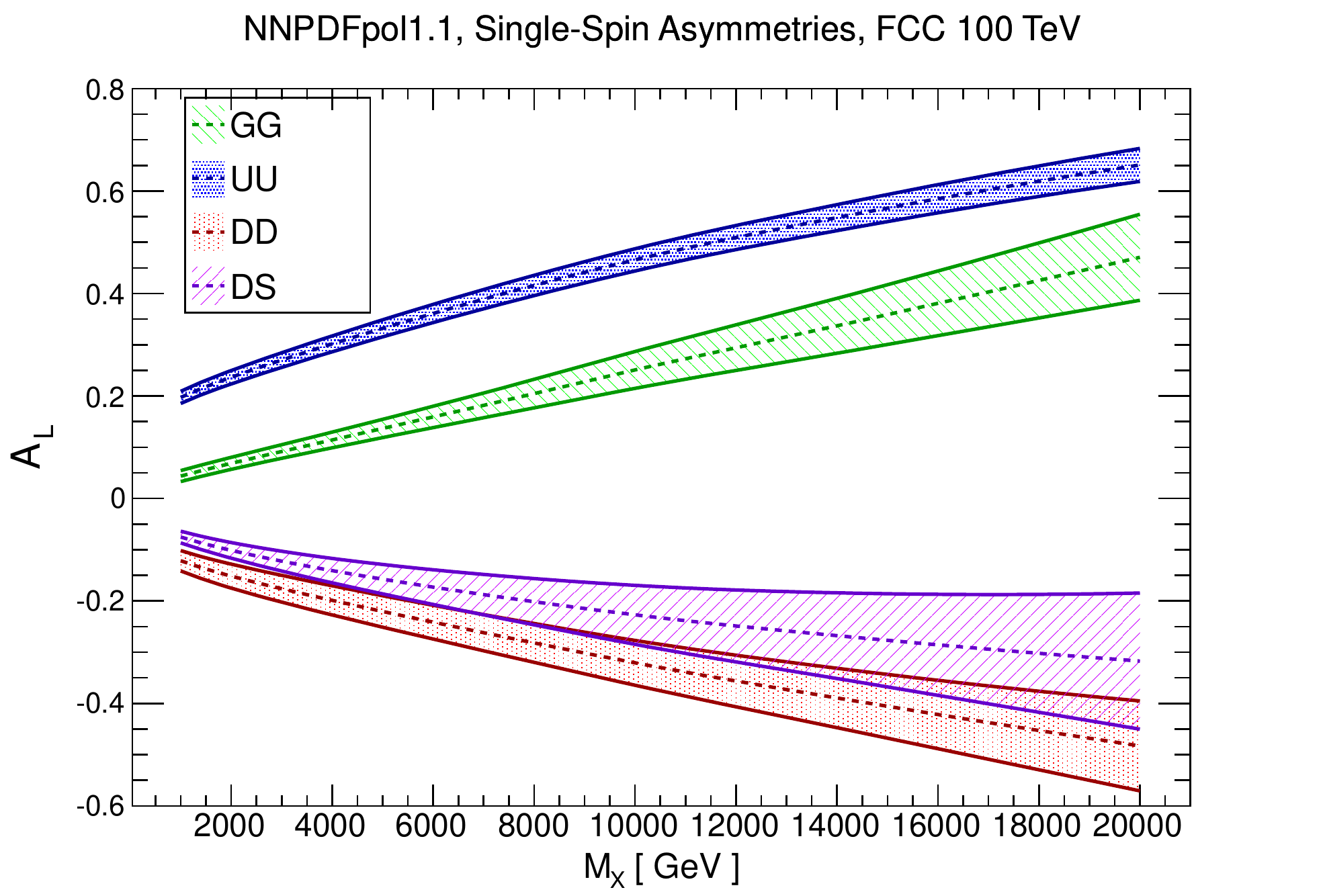}
\caption{\small Summary of the single-spin asymmetries $A_{L}$ for a variety of
initial state partonic combinations as a function of the invariant mass
of the produced final state $m_X$ at the \mbox{LHC 14 TeV} (left panel) and 
at an FCC 100 TeV (right panel).
The asymmetries have been obtained using NNPDFpol1.1/NNPDF2.3.
}
\label{fig:AL_summary}
\end{center}
\end{figure}

Results for the single-spin asymmetries in the
$gg$, $uu$, $dd$ and $ds$ partonic sub-channels
for LHC 14 TeV and FCC 100 TeV
are summarized in Figure~\ref{fig:AL_summary}.
It is apparent that the property which we have
discussed above, namely that if the final state mass
range is suitably scaled, the qualitative features of the
spin asymmetries are the same at center-of-mass energies of
14~TeV and 100~TeV.
The most striking observed property is that different
partonic sub-channels lead to very different asymmetries.
In this particular case, just a measurement of the sign of the asymmetry would
indicate which are the dominant partonic initial states, and measurements
of $A_{\rm L}$ with a few percent experimental uncertainty would even
distinguish between $gg$ and $qq$ initiated final states.

\begin{figure}
\begin{center}
\epsfig{width=0.32\textwidth,figure=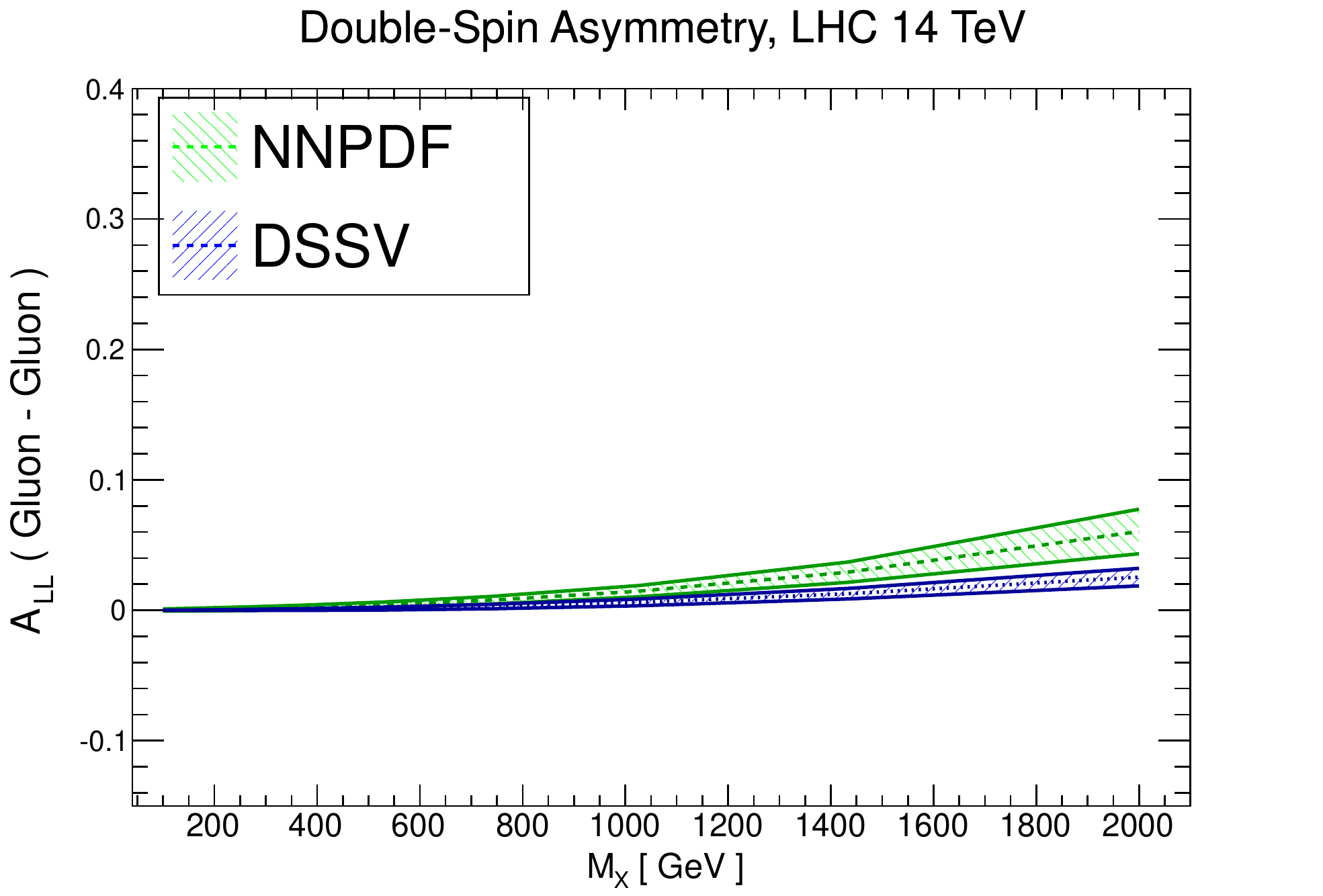}
\epsfig{width=0.32\textwidth,figure=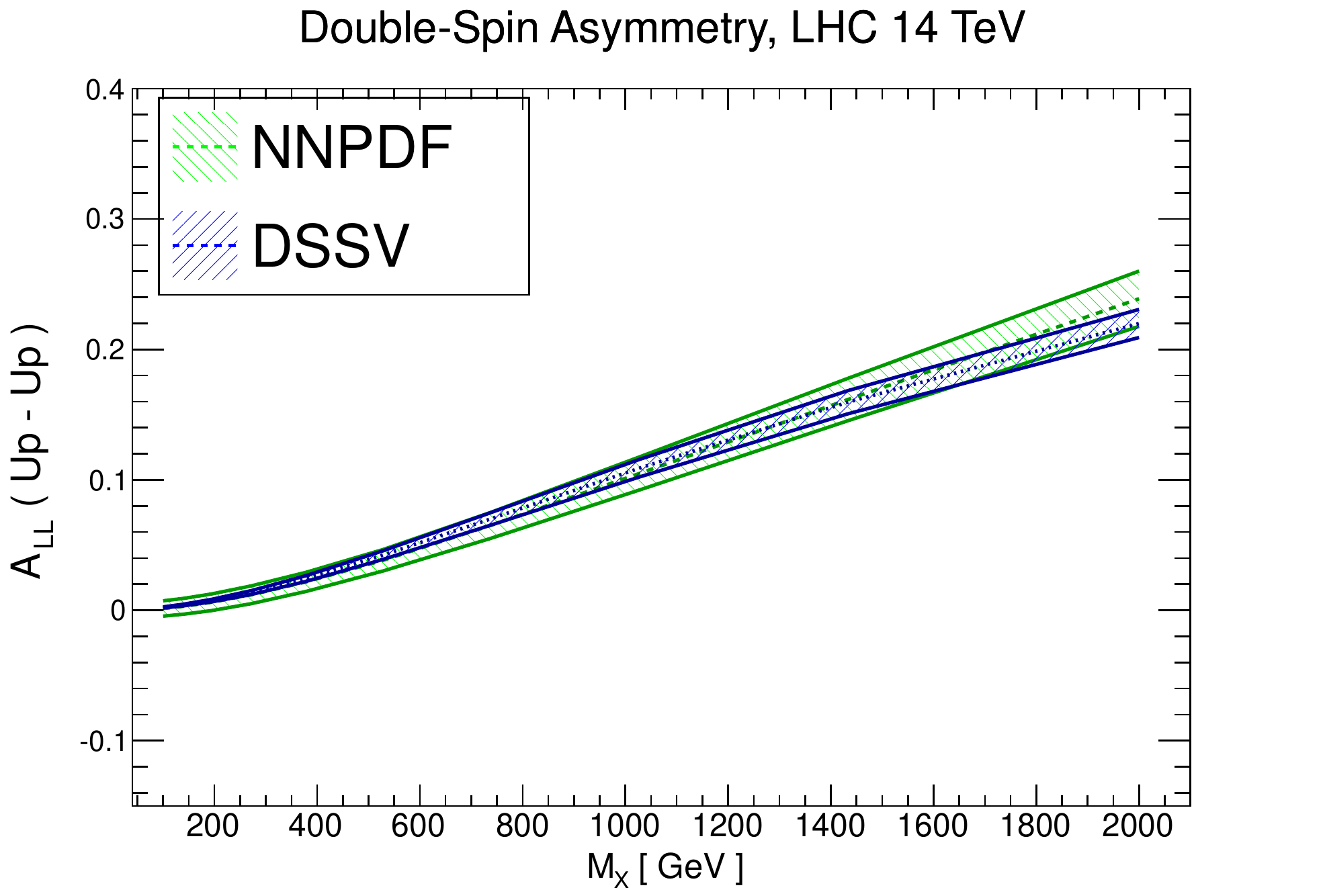}
\epsfig{width=0.32\textwidth,figure=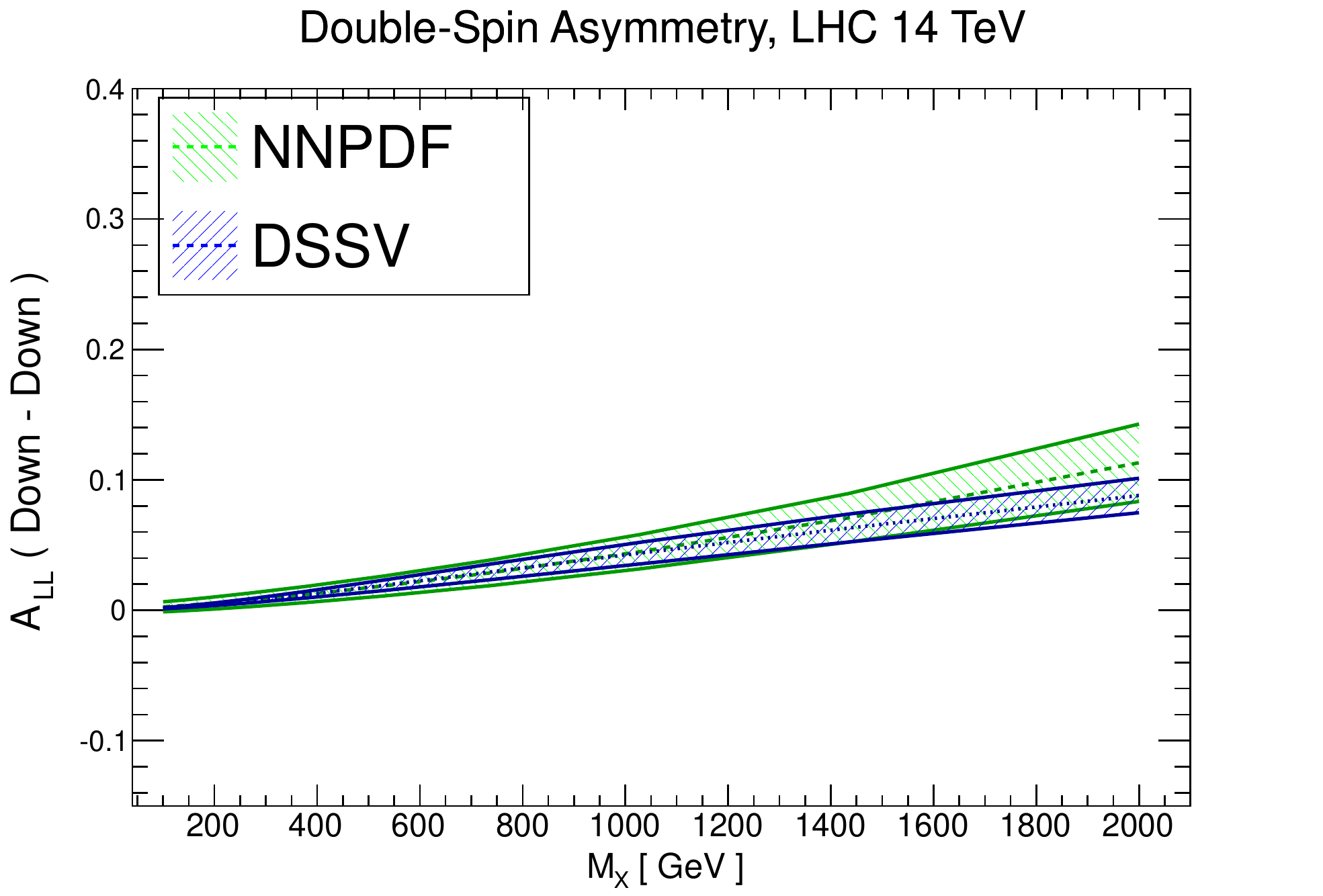}\\ \vspace{.4cm}
\epsfig{width=0.32\textwidth,figure=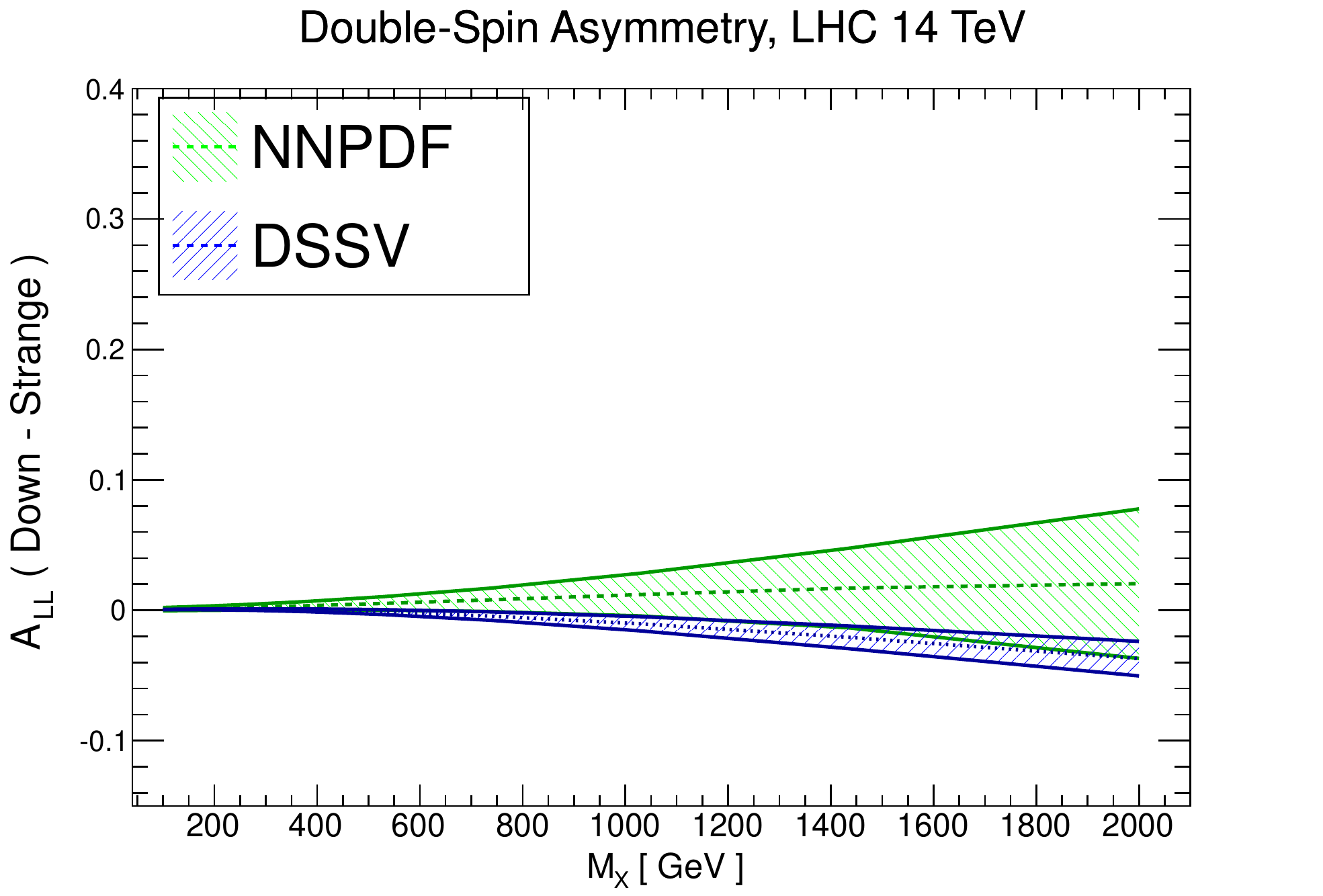}
\epsfig{width=0.32\textwidth,figure=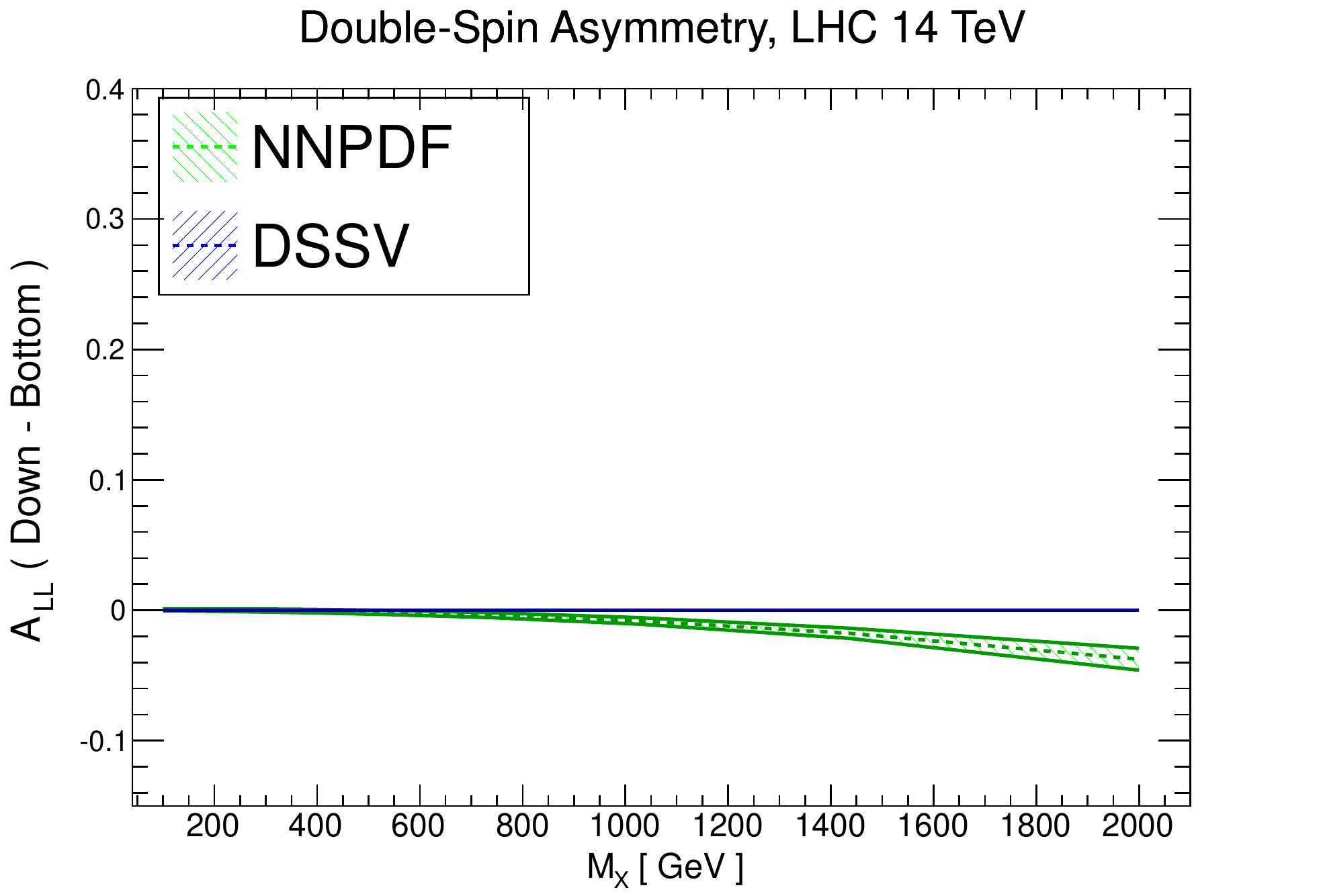}
\epsfig{width=0.32\textwidth,figure=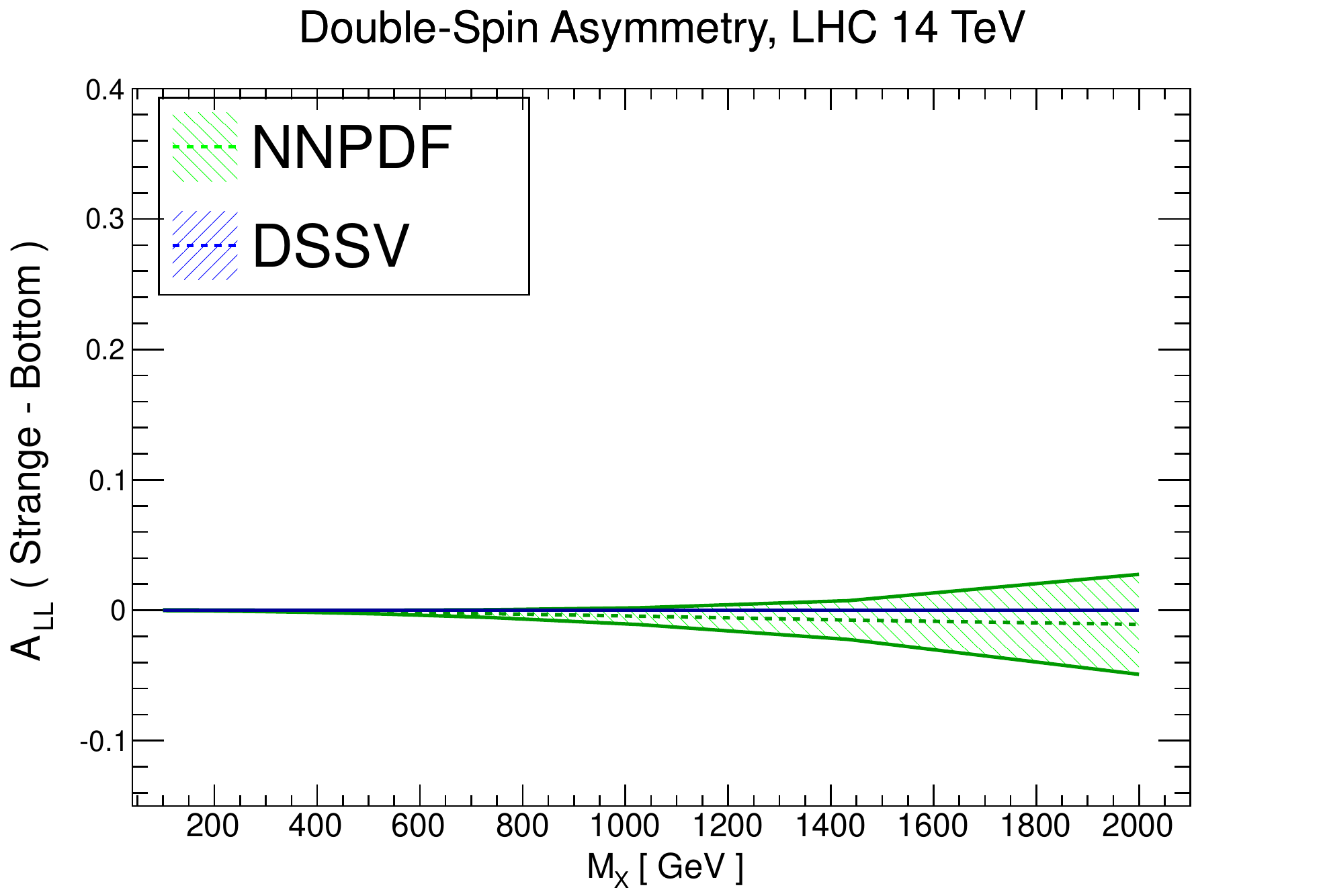}
\caption{\small The double-spin asymmetry $A_{LL}$ at the
PDF level, Eq.~(\ref{eq:al}), at LHC 14 TeV for various initial-state
partonic combinations, comparing the results obtained using NNPDFpol1.1/NNPDF2.3
with those obtained using DSSV/MRST.}
\label{fig:ALL_LHC14}
\end{center}
\end{figure}

Double-spin asymmetries are experimentally more challenging since their absolute values are
smaller.
The reason for this is because they involve the convolution of two polarized PDFs in the numerator,
instead of just one as for single-spin asymmetries.
The results for various partonic channels are summarized in Figure~\ref{fig:ALL_LHC14}.
Again reasonable agreement between NNPDFpol1.1 and DSSV08 is found. 
Asymmetries involving quarks are larger than those involving gluons, reflecting
that the $\Delta q$ densities are larger than the $\Delta g$ one at large-$x$, as shown in
Figure~\ref{fig:pdfcomp}.
Moreover, large final state masses are required to yield
spin asymmetries that are larger than a few percent.

\begin{figure}
\begin{center}
\epsfig{width=0.49\textwidth,figure=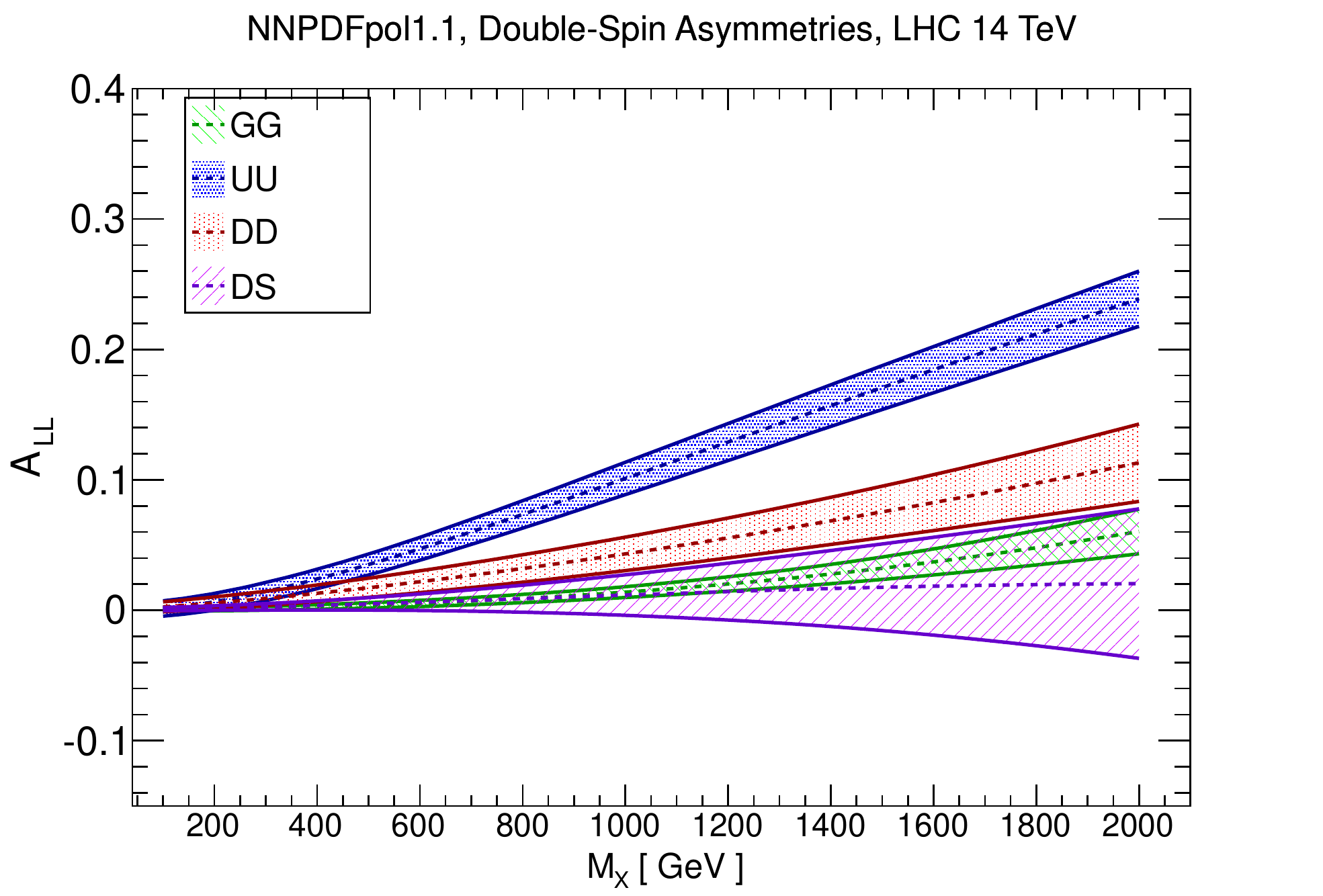}
\epsfig{width=0.49\textwidth,figure=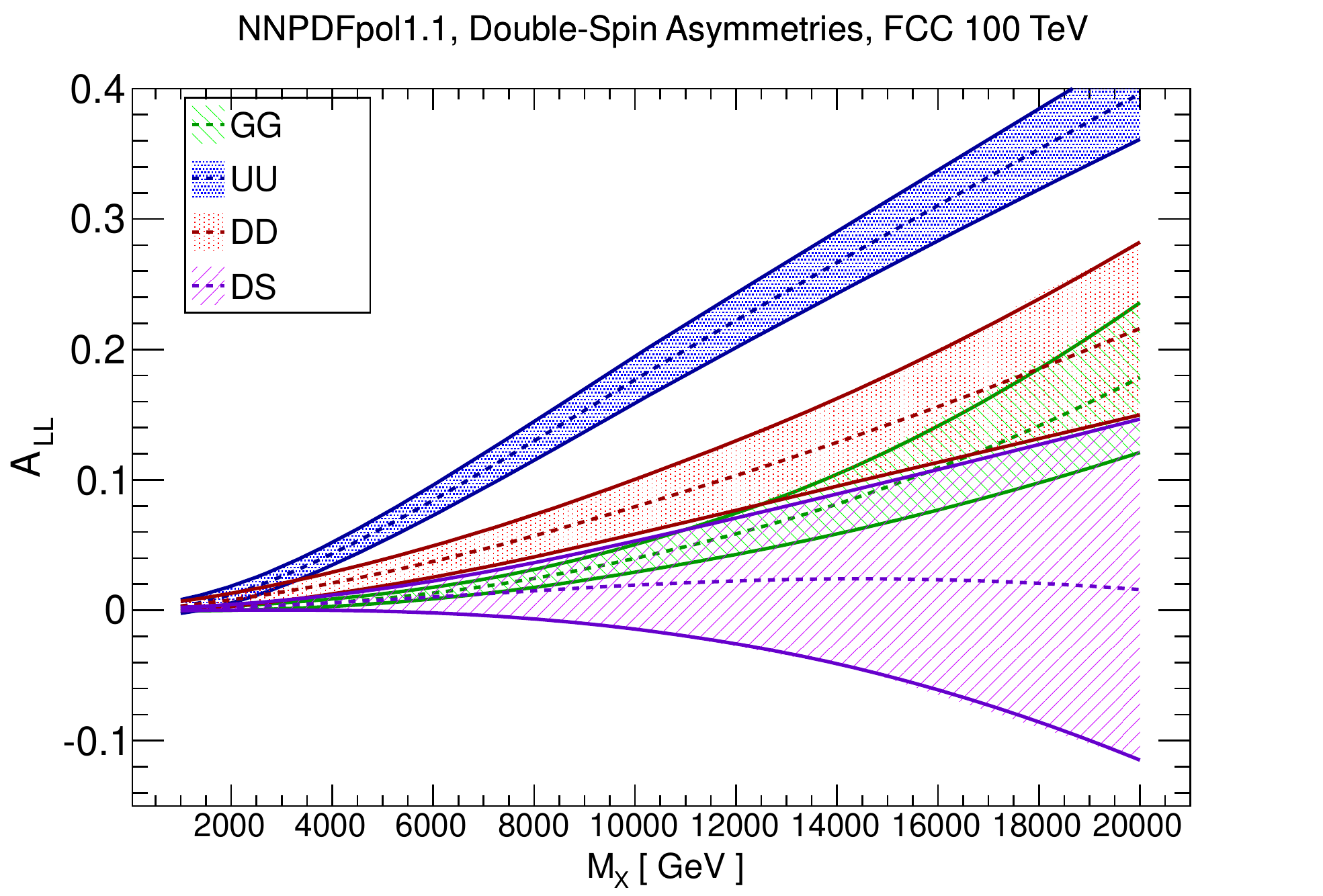}
\caption{\small Summary of the double-spin asymmetries $A_{LL}$ for a variety of
initial state partonic combinations as a function of the invariant mass
of the produced finale state $m_X$ at the \mbox{LHC 14 TeV} (left panel) and 
at an FCC 100 TeV (right panel).
The asymmetries have been obtained using NNPDFpol1.1.
}
\label{fig:ALL_summary}
\end{center}
\end{figure}

The final comparison is provided by double-spin asymmetry calculations
for the $gg$, $uu$, $dd$ and $ds$ partonic sub-channels
for LHC 14 TeV and FCC 100 TeV and we collect the results
in Figure~\ref{fig:ALL_summary}.
The PDF uncertainties are found very large,
since there is far less experimental
information in the determinations of the $\Delta q_i$ densities
than in the unpolarized case.
However, if the measurement of a vanishing double-spin asymmetry
could be performed, this would be a valuable
piece of information since it would exclude that the final
state is dominantly produced from $uu$ scattering. Moreover, double-spin asymmetry
measurements could nevertheless be used to verify the results obtained from
single-spin asymmetries.

After this discussion at the PDF level only, in the next section we will present
predictions for hadron-level asymmetries in various scenarios for
BSM monotop production.
From now on we will neglect for clarity the polarized PDF uncertainties.
It has already been shown in this section that they are large, however,
the availability of a polarized hadron collider would also provide
a large set of polarized PDF-sensitive measurements that should substantially
reduce these uncertainties.
In addition, in the short term, additional constraints from a variety
of polarized measurements from fixed target and collider  experiments
like HERMES, COMPASS and RHIC will allow to further pin down the polarized
PDFs.
In the medium term, important constraints on polarized PDFs could also be provided 
by a Electron-Ion 
Collider (EIC)~\cite{Boer:2011fh,Ball:2013tyh,Aschenauer:2013iia,Aschenauer:2012ve}, currently
under study.

\section{Pinning down monotop production dynamics with polarized beams}
\label{sec:monotops}

In order to illustrate the power of spin asymmetries for the
characterization of new physics, we focus in this work
on one generic BSM signature, dubbed
monotop, that has recently been proposed~\cite{Andrea:2011ws, Agram:2013wda}.
The monotop signature is characterized by a
top quark produced singly in association with
missing energy and without any additional particle.
The choice of such a
process to illustrate the usefulness of polarized proton-proton collisions for
physics beyond the Standard Model is driven by several
considerations.

First of all, the sector of the top quark is widely believed to
be one of the key candidates for coupling in an enhanced way to new physics
particles due to the vicinity of the top mass to the electroweak scale.
Second, monotop production 
is negligible in the Standard Model
where the top quark is produced in association with a $Z$-boson 
and no extra jet.
This process is indeed loop-induced, GIM-suppressed and further reduced by the
branching ratio of the $Z$-boson to neutrinos. 
This ensures that a monotop
observation at the LHC (or at an FCC) can be safely thought
of as a clear tell-tale sign of new physics. 
Finally, there exists a wide variety of new physics theories
that lead to the same final-state monotop signature 
(see, \textit{e.g.}, Refs.~\cite{Berger:1999zt,
Berger:2000zk, Desai:2010sq, Davoudiasl:2010am,Davoudiasl:2011fj, Kamenik:2011nb, Andrea:2011ws,
Fuks:2012im, Alvarez:2013jqa, Agram:2013wda}).
This process therefore offers a
good way to illustrate how single-spin and double-spin asymmetries can provide
a unique handle to extract information on
the underlying theory realized in Nature,
should monotop production be observed.

\subsection{Monotop production in the RPV MSSM}
\label{sec:rpv}

We begin by considering the MSSM
after supplementing its $R$-parity conserving
superpotential by one single RPV
operator, the so-called UDD term.
As it will be shown below, this simple setup includes three
distinct monotop production mechanisms hardly distinguishable in unpolarized
proton-proton collisions, apart from the differences in
total rates that are however
dependent on unknown couplings.
This contrasts with the situation when additional polarized
observables are available since in this case,
discriminating between the
different initial states becomes possible.

\begin{figure}
\centering
  \vspace{-1.0cm}
  \epsfig{width=1\columnwidth,figure=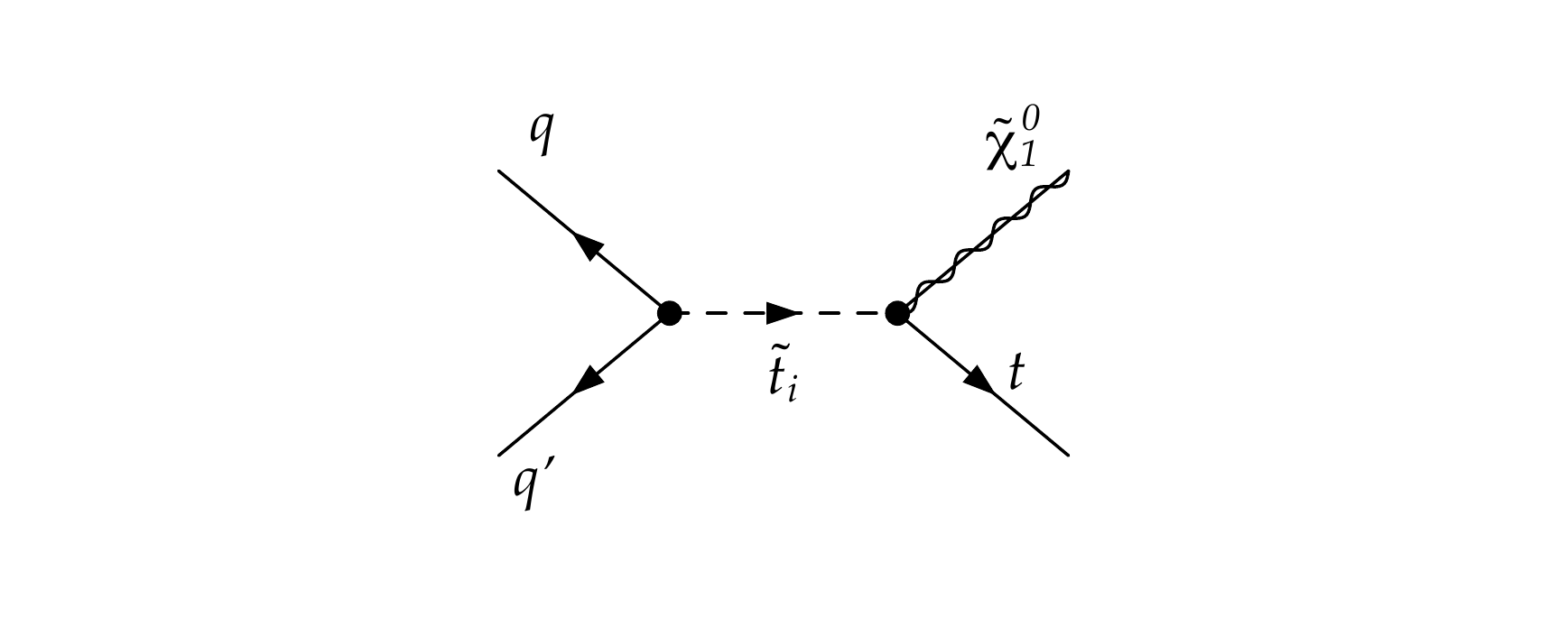}
  \vspace{-1.4cm}
  \caption{\small \label{fig:rpvdiag}Feynman diagram associated with
  RPV monotop
  production. The neutralino $\tilde\chi_1^0$ is assumed long-lived so that
  it decays outside the detector, effectively giving rise to missing
transverse energy.
  Non-resonant subprocesses have here been neglected since
they are small compared to resonant production.}
\end{figure}

We model the supersymmetric
interactions among the matter sector
by adding to the $R$-parity conserving MSSM superpotential
$W_{\rm MSSM}$ the RPV $UDD$ operator,
\be
  W = W_{\rm MSSM} + \frac{1}{2}\lambda_{ijk}^{''} U_R^{i}D_R^{j}D_R^{k} \ ,
\label{eq:rpvsuperW}\ee
where $U_R$ and  $D_R$ are the chiral superfields associated with the
up-type and down-type right-handed (s)quark supermultiplets, the color indices
are implicit for clarity and the flavor indices being explicitly indicated.
Monotop production is induced by non-vanishing $\lambda_{3jk}^{''}$ couplings
together with enforcing the lightest neutralino to be long-lived, a setup
almost unconstrained by experimental data~\cite{Barbier:2004ez}.
If at least one of these $\lambda^{''}$ couplings is
non-vanishing, top squarks of mass $m_{\tilde t}$
can be resonantly produced from the scattering of two
down-type antiquarks of different flavors and further decay into a top quark and
a lightest neutralino (see Figure~\ref{fig:rpvdiag}) which,
if lighter than the top quark, is long-lived enough to escape detection and
gives rise to missing transverse energy in a detector~\cite{Allanach:1999bf}.
The same final state
could also be produced through $t/u$-channel down-type squark exchanges allowed
by the interactions driven by the superpotential of Eq.~\eqref{eq:rpvsuperW}. We
however neglect, in the following, these non-resonant contributions with respect
to the resonant $s$-channel diagram, that we have explicitly
verified to be largely dominant.

In the RPV supersymmetric framework described above, the fully polarized
partonic cross-section for monotop production from a
$\bar{q}\bar{q}'$ initial-state is given by
\be\label{eq:rpvpol}
  \hat\sigma^{h_1h_2}_{RPV} (\bar q_j \bar q_k \to t \tilde\chi_1^0)=
   \frac{(1-h_1)(1-h_2) \pi \big| \lambda^{\prime\prime}_{3jk}
     \sin\theta_{\tilde t} \big|^2}{6} \ \
      \textrm{BR}\Big(\tilde t \to t \tilde \chi^0_1\Big)\ \
   \delta\big(\hat s - m_{\tilde t}^2\big) \ ,
\ee
where $\hat s$ denotes the partonic center-of-mass energy
and $h_1$ and $h_2$ the helicities of the initial
antiquarks.
Results for the charge-conjugate process
can be obtained by replacing $h_i \to -h_i$.
Moreover, we have assumed that only one of the two stop mass-eigenstates
is light enough to significantly contribute to the cross-section and
kept the associated dependence on the stop mixing angle $\theta_{\tilde t}$ explicit.
Finally, we have adopted the narrow-width approximation to model the
resonant behavior of the squared matrix element by a Breit-Wigner lineshape.
Although non-general, such an approximation holds when the width of the
resonance is small with respect to its mass, which allows one to neglect
off-shell effects, when the resonance decays into much lighter particles and
when its mass is much smaller than the
center-of-mass energy, which avoids important distortions of the Breit-Wigner
lineshape~\cite{Berdine:2007uv}.

Because of
the symmetry properties of the RPV superpotential of Eq.~\eqref{eq:rpvsuperW},
the couplings of quarks or antiquarks of the same flavor 
to the stop $\tilde{t}$ vanish so that only three different
flavor combination can yield a non-zero cross-section, namely the
$d\,s+\bar d\,\bar s$, $d\,b+\bar d\, \bar b$ and $s\, b+\bar s\, \bar b$ 
initial states.
Here we are implicitly summing
over both monotop and anti-monotop production, while later in the section,
we will explore the potential of tagging the charge of the
final-state monotop.
The parton luminosities that contribute to the unpolarized
cross-section, the single- and the double-spin asymmetries in the various
different scenarios for monotop production that are discussed in this
paper are summarized in Table~\ref{tab:pdflumis}.

\renewcommand{\arraystretch}{1.2}
\begin{table}
\begin{center}
\begin{tabular}{c||c|c|c}
\hline
Scenario  & $\sigma_0$ &  $\sigma_L$ &   $\sigma_{LL}$  \\
\hline
\hline
\multirow{3}{*}{MSSM RPV} &  $d\,s+\bar{d}\,\bar{s}$   & 
$d\,\Delta s + s\,\Delta d +
\bar{d}\,\Delta \bar{s} + \bar{s}\,\Delta \bar{d}
 $   &  
$\Delta d \,\Delta s+\Delta \bar{d} \,\Delta \bar{s}$  \\
 &  $d\,b+\bar{d}\,\bar{b}$   &  
$d\,\Delta b + b\,\Delta d +
\bar{d}\,\Delta \bar{b} + \bar{b}\,\Delta \bar{d}
 $ 
  & 
$\Delta d \,\Delta b+\Delta \bar{d}\, \Delta \bar{b}$   \\
 &  $s\,b+\bar{s}\,\bar{b}$   &   
$s\,\Delta b + b\,\Delta s +
\bar{s}\,\Delta \bar{b} + \bar{b}\,\Delta \bar{s}
 $ 
 & 
$\Delta s \,\Delta b+\Delta \bar{s}\, \Delta \bar{b}$   \\
\hline
\multirow{3}{*}{Hylogenesis} &  $d\,d+\bar{d}\,\bar{d}$   & 
$ d\,\Delta d+ \bar{d}\,\Delta\bar{d}$
   &    
$\Delta d \,\Delta d+\Delta \bar{d}\, \Delta\bar{d}$ \\
 &  $s\,s+\bar{s}\,\bar{s}$   & 
$ s\,\Delta s+ \bar{s}\,\Delta\bar{s}$
   &  $\Delta s \,\Delta s+\Delta \bar{s}\, \Delta\bar{s}$  \\
 &  $b\,b+\bar{b}\,\bar{b}$   & 
$ b\,\Delta b+ \bar{b}\,\Delta\bar{b}$
  &   $\Delta b\, \Delta b+\Delta \bar{b}\, \Delta\bar{b}$ \\
\hline
\multirow{2}{*}{$X$-model} &  $g\,u+g\,\bar{u}$   & 
$g \,\Delta u + u\,\Delta g + g \,\Delta \bar{u} + \bar{u}\,\Delta g$ 
    & 
$\Delta g \,\Delta u+\Delta g\,\Delta \bar{u}$   \\
 &  $g\,c+g\,\bar{c}$  & 
$g \,\Delta c + c\,\Delta g + g \,\Delta \bar{c} + \bar{c}\,\Delta g$ 
   & $\Delta g \,\Delta c+\Delta g\,\Delta \bar{c}$    \\
\hline
\end{tabular}
\caption{\small Parton luminosities that contribute to the unpolarized
cross-section, the single- and the double-spin asymmetries in the three
different scenarios for monotop production that are discussed in this
paper.
For each model, the first row corresponds to the dominant production
channel.
In singly-polarized collisions, the second hadron is the one that
we choose to be polarized.
 \label{tab:pdflumis}}
\end{center}
\end{table}
\renewcommand{\arraystretch}{1.0}

\begin{figure}
\centering
  \epsfig{width=.49\columnwidth,figure=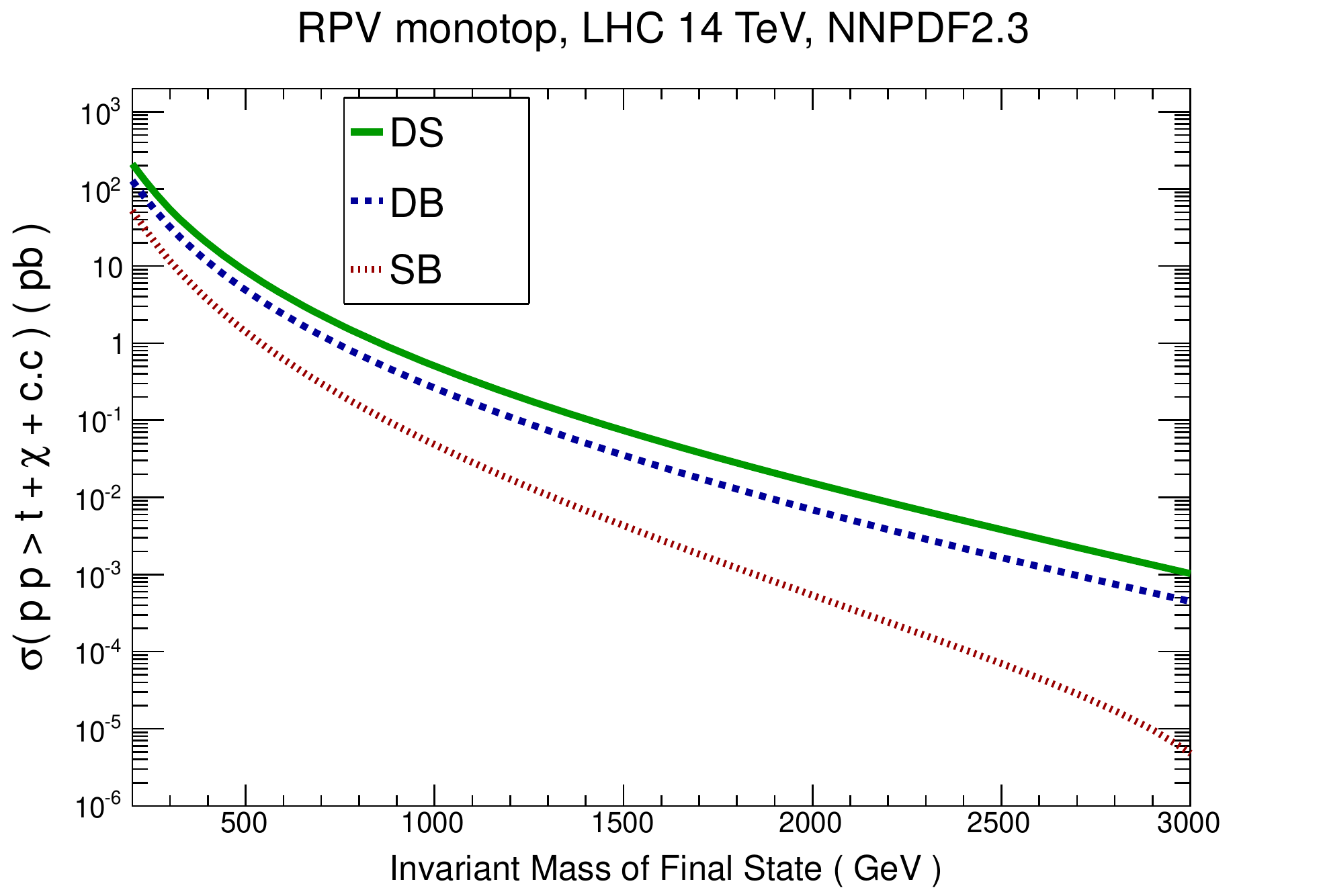}
  \epsfig{width=.49\columnwidth,figure=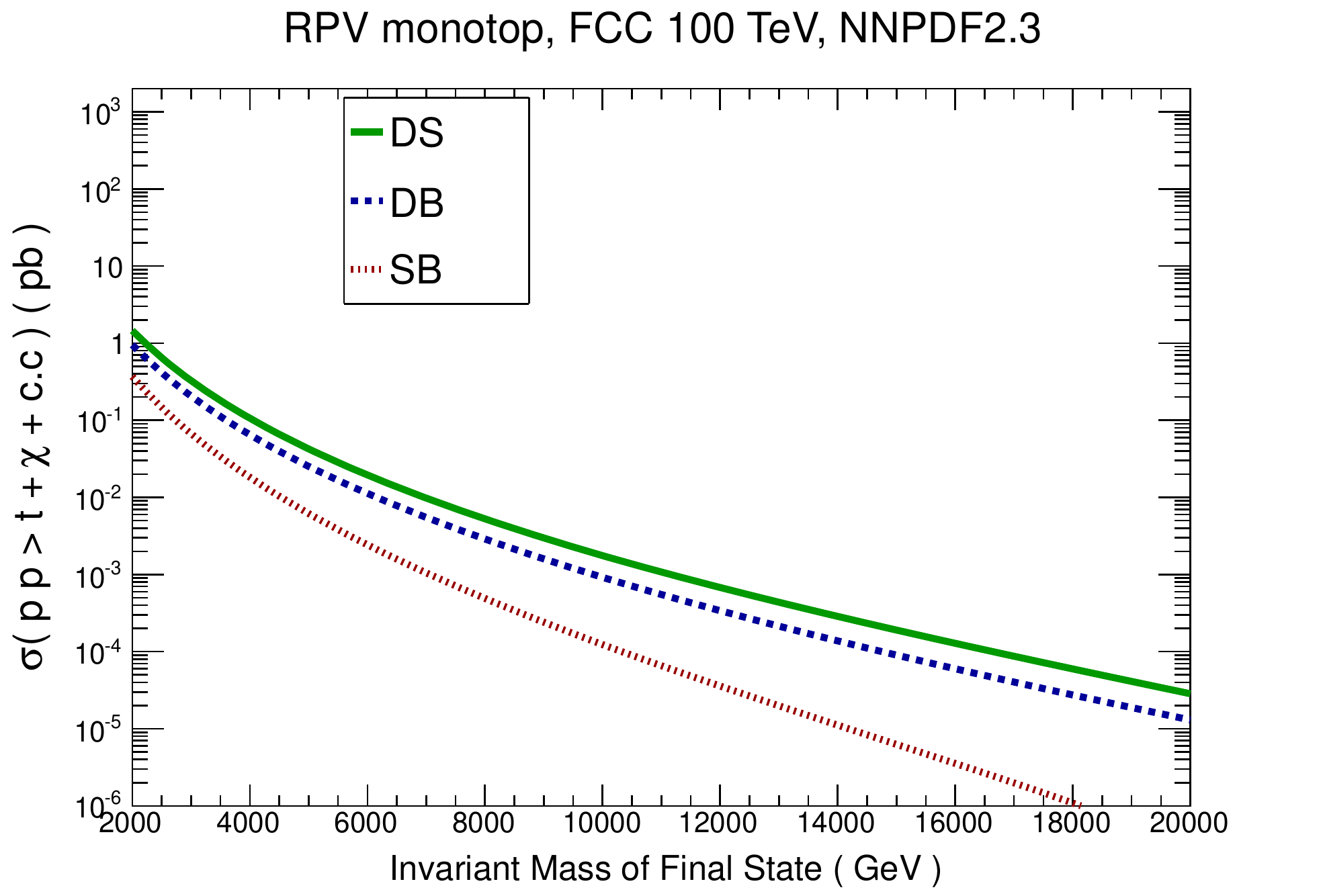}
  \caption{\small \label{fig:rpvxsection}RPV monotop production total cross
  sections at the LHC 14 TeV (left panel) and at an FCC 100 TeV (right panel)
  as function of the invariant mass of the final state.
  We fix the stop mixing angle to $\pi/4$,
  consider the branching ratio $\text{BR}(\tilde t\to t \tilde\chi_1^0)=1$
  and address three distinct benchmark scenarios
  where one single RPV coupling is non-vanishing at a time:
  $\lambda^{''}_{312}$ (green),  $\lambda^{''}_{313}$ (blue) and
  $\lambda^{''}_{323}$ (red). Cross-sections have been obtained using
the NNPDF2.3 unpolarized parton set.}
\end{figure}

In Figure~\ref{fig:rpvxsection}, we present total cross-sections for RPV monotop
production at the \mbox{LHC 14 TeV} (left panel of the figure)
and at an \mbox{FCC 100 TeV} (right panel of the figure) as function of the
mass of the lightest top squark. We compute our results
by making use of Eq.~\eqref{eq:sigunpol} with the NNPDF2.3 set of parton
densities, and for the sake of the example, we consider maximal stop
mixing ($\theta_{\tilde t} = \pi/4$), the branching ratio of the stop resonance
into a monotop
state equal to unity ($\textrm{BR}(\tilde t \to t \tilde \chi^0_1) = 1$) and
all the three possible different initial states. We however
assume that only one of the three RPV couplings is non-zero at a time and
that its value is fixed to $\lambda^{''}=0.2$. With a quadratic dependence on
the $\lambda''$-parameters, RPV monotop production could be in principle expected
both at the LHC and at an FCC. However, characterizing which partonic
initial state would be (dominantly) responsible for the possible observation of an
excess is far more complicated than measuring a total cross-section. 
The standard approach would then 
be to probe differential distributions sensible, \textit{e.g.},
to the presence of valence or sea quarks in the initial state.
However, in the rest of this section we focus on a complementary approach
to characterize the initial state of
monotop production by means of spin asymmetry measurements
in polarized $pp$ collisions.

\begin{figure}
\centering
  \epsfig{width=.49\columnwidth,figure=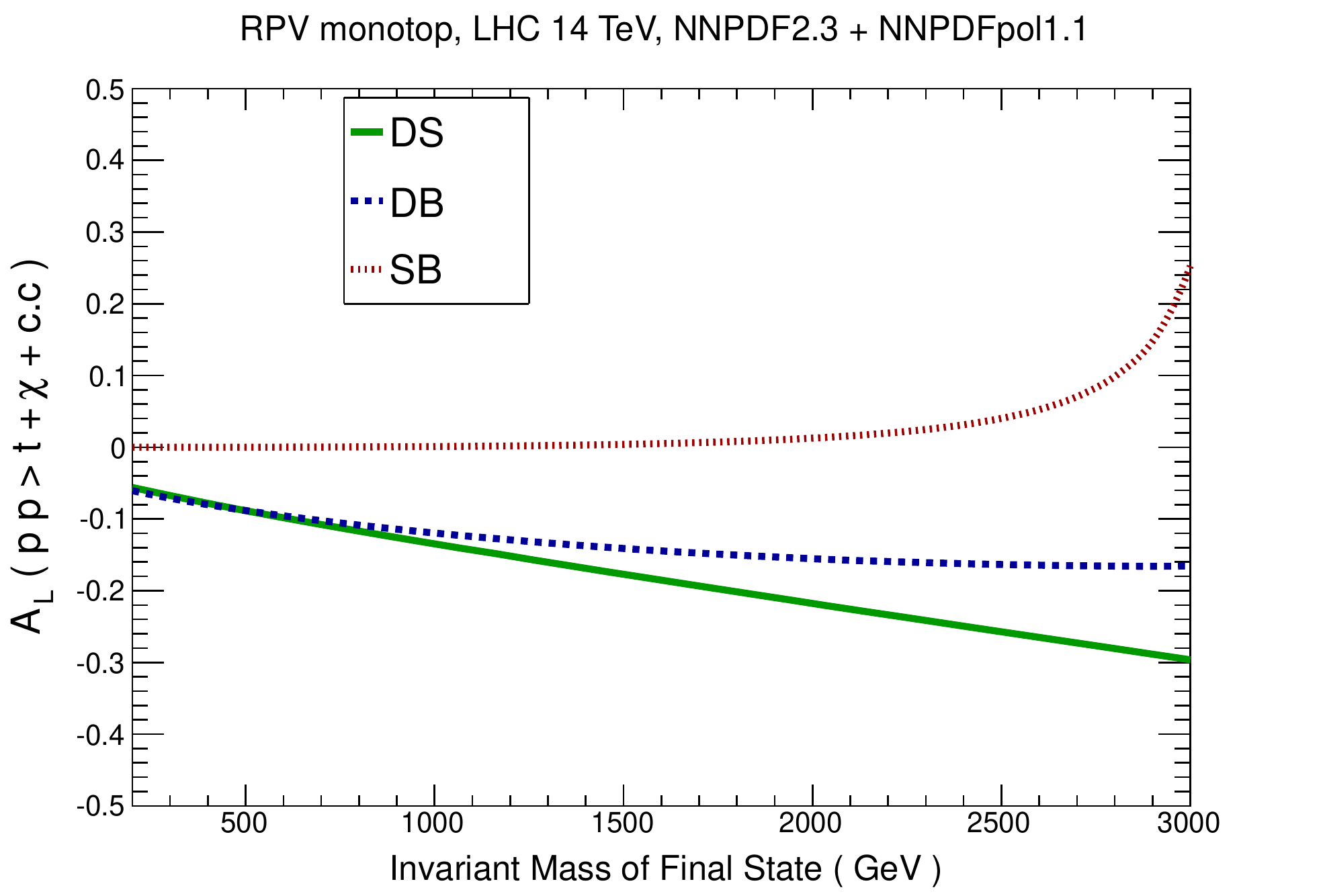}
  \epsfig{width=.49\columnwidth,figure=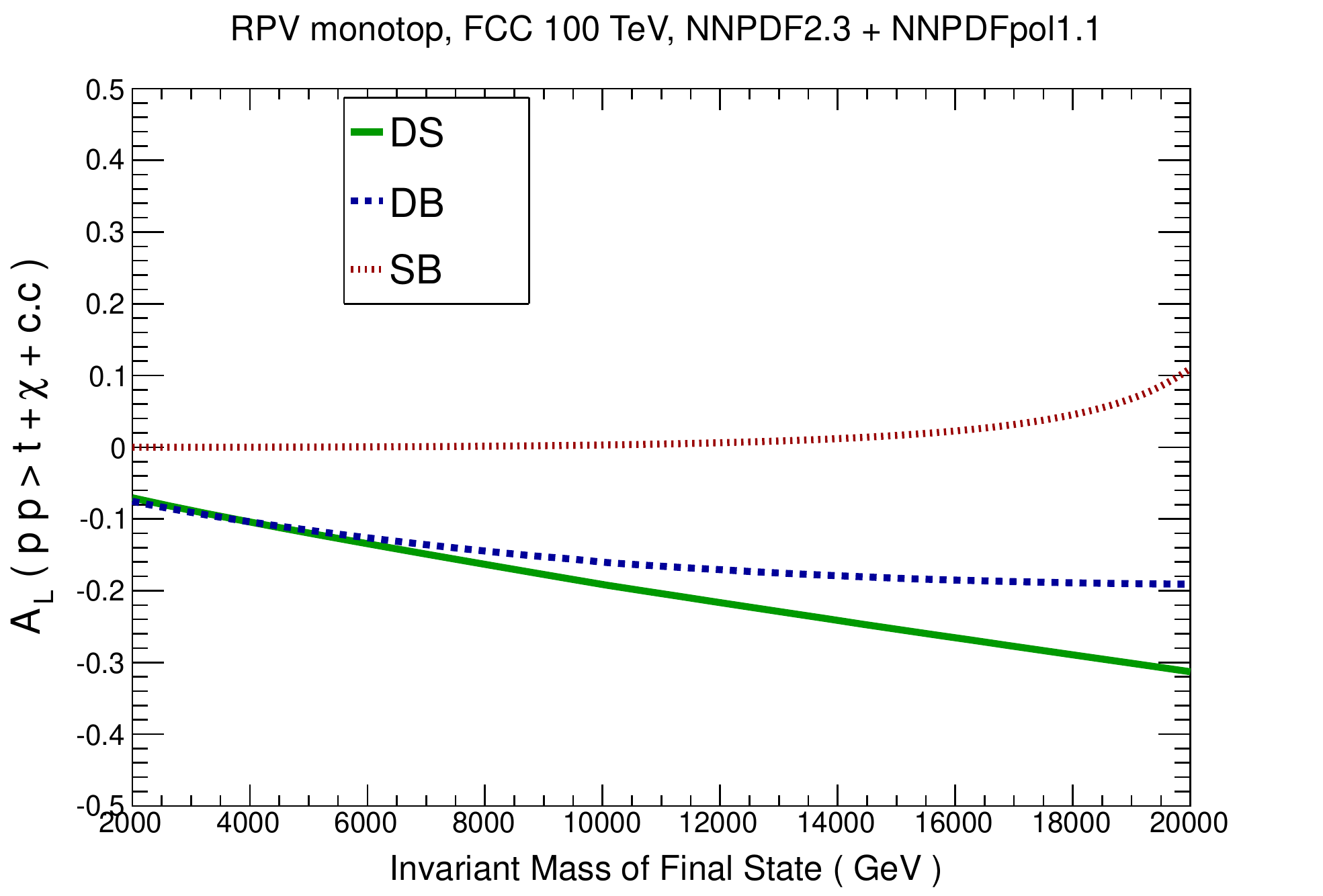}\\ \vspace{.3cm}
  \epsfig{width=.49\columnwidth,figure=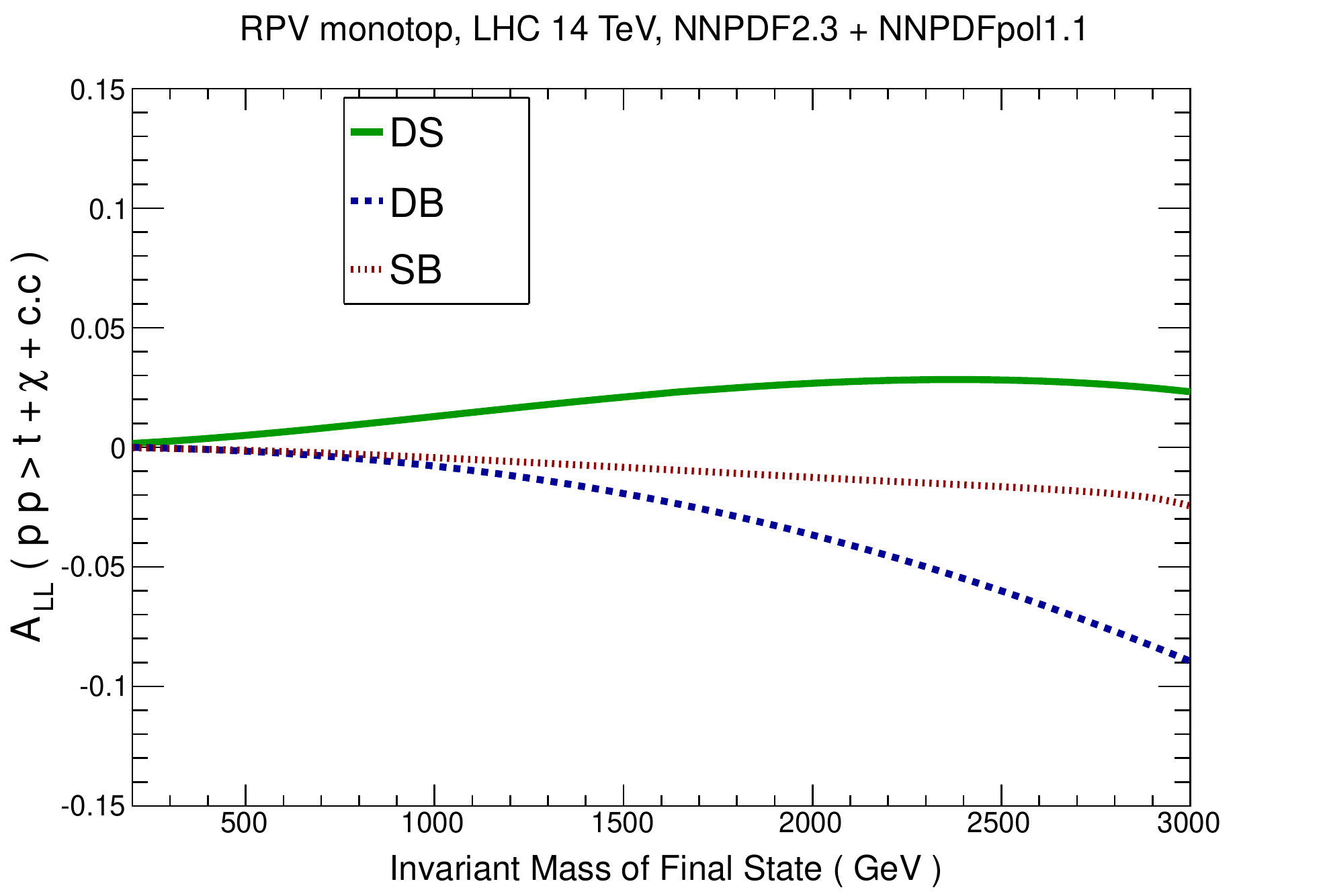}
  \epsfig{width=.49\columnwidth,figure=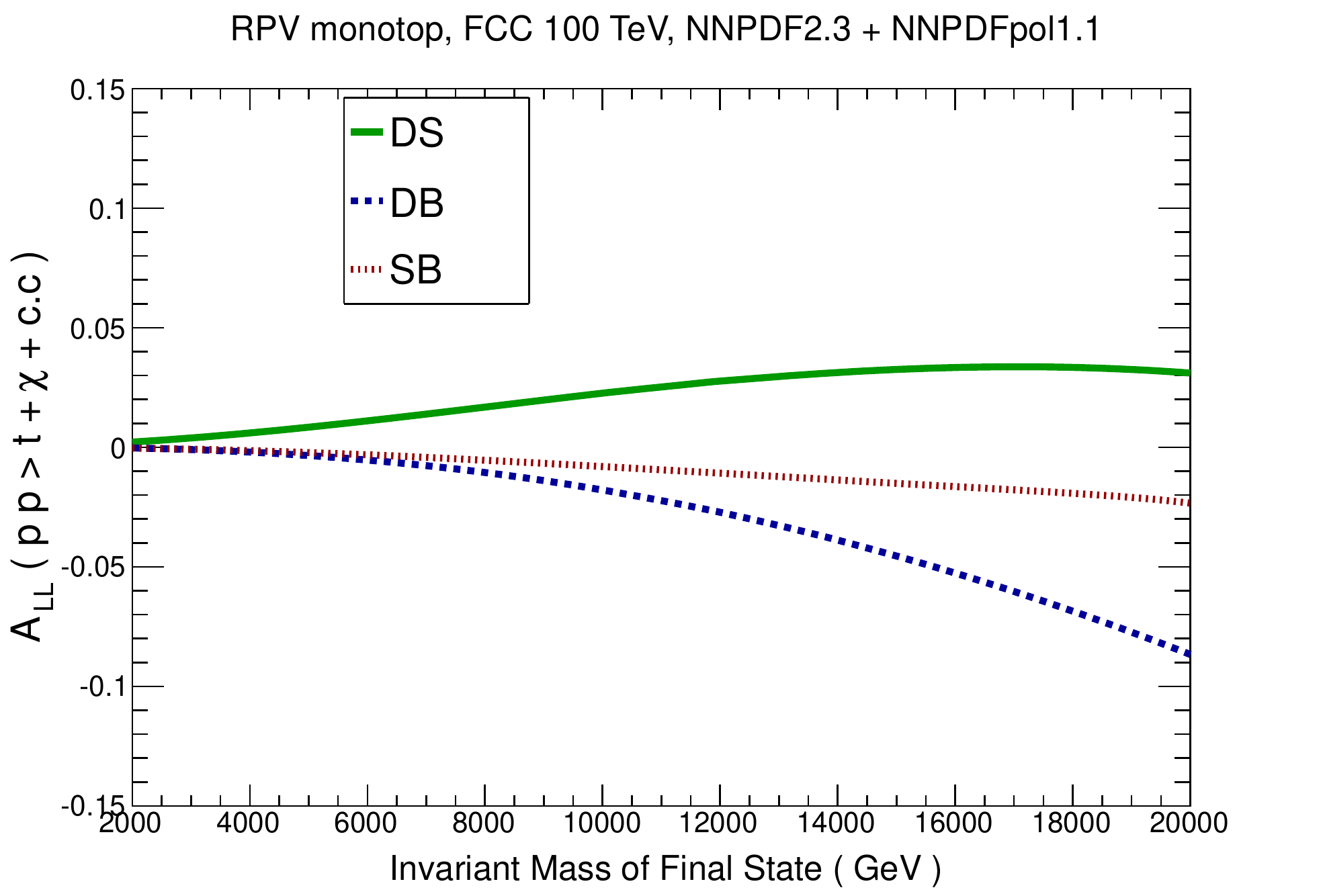}
  \caption{\small \label{fig:rpvasym}Single-spin (upper panel) and double-spin (lower panel)
  asymmetries for RPV monotop production at the LHC 14 TeV (left panel) and at an
  FCC 100 TeV (right panel) as function of the stop (or monotop) mass.
  We fix the stop mixing angle to $\pi/4$,
  consider the branching ratio $\text{BR}(\tilde t\to t \tilde\chi_1^0)=1$
  and address three distinct benchmark scenarios
  where one single RPV coupling is non-vanishing at a time:
  $\lambda^{''}_{312}$ (green),  $\lambda^{''}_{313}$ (blue) and
  $\lambda^{''}_{323}$ (red). Asymmetries have been obtained using NNPDFpol1.1 and NNPDF2.3.
}
\end{figure}

In the RPV context, there is only one single combination of quark helicities
that gives rise to a monotop final state, as indicated in Eq.~\eqref{eq:rpvpol}. 
Consequently,
partonic spin asymmetries turn out to be equal to $\pm 1$ and
hadronic asymmetries reduce to ratios of partonic luminosities. Therefore,
in the approximation in which there is a single dominant coupling
$\lambda''$, hadron-level spin-asymmetries can be expressed in terms of a ratio
of linear combinations of polarized and unpolarized PDFs and the results of
Section~\ref{sec:lumipol} hold.
For instance, for the case of monotop production in the dominant
channel 
$\bar{d}\,\bar{s}$+$d\,s$ we have
\be
A_{L}^{\bar{d}\bar{s}+ds} = \frac{\mathcal{L}^{L}_{ds}-\mathcal{L}^{L}_{\bar{d}\bar{s}}}{\mathcal{L}_{ds}+\mathcal{L}_{\bar{d}\bar{s}}} \qquad\text{and}\qquad
A_{LL}^{\bar{d}\bar{s}+ds} = \frac{\mathcal{L}^{LL}_{ds}+\mathcal{L}^{LL}_{\bar{d}\bar{s}}}{\mathcal{L}_{ds}+\mathcal{L}_{\bar{d}\bar{s}}} \, ,
\ee
and likewise for other initial states.
We collect the results for the relevant
channels in Figure~\ref{fig:rpvasym} for both the LHC 14 TeV (left panel)
and the \mbox{FCC 100 TeV} (right panel), after summing over both monotop and
anti-monotop production modes, and show single-spin (upper row of the figure)
and double-spin (lower row of the figure) asymmetries.

It is clear from Figure~\ref{fig:rpvasym} that polarized asymmetries can
be sizable, and moreover depend strongly on the partonic initial state.
For instance, at the LHC 14 TeV and $m_X=3$~TeV, $A_L$ varies from
$20\%$ for the $sb$ initial state to $-30\%$ for the $ds$ combination.
The different behaviors of $A_{LL}$ and $A_L$ for the same partonic
initial state has also discrimination power.
Therefore, it can be seen that the availability of polarized
beams at high energy hadron colliders allows
to disentangle the different possible
scenarios leading to monotop production, especially for large 
final-state masses,
where the polarized asymmetries are larger.

\subsection{Other scenarios for monotop production}

In addition to the RPV MSSM scenario,
several other models predict monotop production at 
hadron colliders.
Therefore, in the event of observation of the monotop signature,
determining which is the correct underlying 
 model will be a difficult task.
In particular, even disentangling a resonant monotop
production from a non-resonant one might be non-trivial due to detector effects
distorting typical resonant shapes expected, for instance, in
the missing energy spectrum~\cite{Agram:2013wda}.
In Section~\ref{sec:rpv}, we have investigated monotop production in the context
of RPV supersymmetry and have shown how spin asymmetries could help
characterizing the type of RPV interactions relevant for the production of a
monotop state. We now investigate two additional scenarios predicting the production
of a top quark in association with missing energy, and illustrate the 
strengths 
of measuring spin asymmetries in polarized collisions
in order to obtain information on the underlying model.

\begin{figure}
\centering
  \vspace{-1.0cm}
  \epsfig{width=0.40\columnwidth,figure=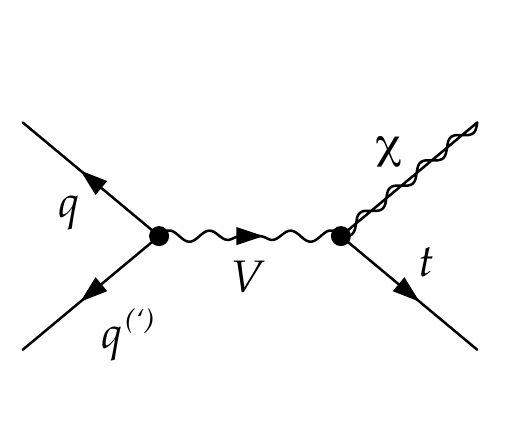}
  \vspace{-0.7cm}
  \caption{\small \label{fig:otherdiag1}Feynman diagram associated with Hylogenesis
  monotop production.}
\end{figure}

We first focus on the so-called Hylogenesis models for dark matter where
a monotop state can be produced from the decay of a heavy
vector resonance $V_\mu$ of mass $m_V$ that couples to down-type
quarks~\cite{Davoudiasl:2010am,Davoudiasl:2011fj}.
The leading order Feynman diagram for monotop production in this scenario
is shown in Figure~\ref{fig:otherdiag1}. 
The heavy vector resonance $V_\mu$ 
 decays into an associated
pair comprised of a top quark and a spin-1/2 dark matter particle, carrying
missing energy, that we generically denote
by $\chi$. We further describe the couplings of down-type quarks
to the colored resonance $V_\mu$ with charge $\pm 2/3$ 
 by the Lagrangian\footnote{The Lagrangian
choice is not unique and we focus on one particular example among others
that induces large differences compared to the RPV case.}
\be
  {\cal L}_{\rm hylo} = \frac12 \kappa_{ij}\ \bar d^c_i \gamma^\mu d_j V_\mu 
 + \text{h.c.} \ ,
\label{eq:Lhylo}\ee
where again a sum over color indices is understood, $i$ and $j$ are flavor indices and
$\kappa_{ij}$ denotes the $3\times 3$ (symmetric) matrix of interaction strengths in
flavor space. 
As in Eq.~\eqref{eq:rpvpol} for the RPV case,
we will rewrite any dependence on the interactions of the
dark matter state $\chi$ in terms of the branching ratio of $V_\mu$ to
a monotop final state, so that the corresponding Lagrangian contributions
are unnecessary and have been omitted.
The constraints on the parameters of this scenario from collider and flavor physics
have been studied in Ref.~\cite{Kim:2013ivd}, which shows that they are model 
dependent and can be easily avoided, 
specially in the case of third generation quarks. 
 In the rest of this section, we
restrict ourselves to the dominant $dd$+$\bar d \bar d$ production channel,
again summing over monotop and anti-monotop production.
As stated above, a discussion on the information that can be obtained by tagging
the charge of the final-state top quark, and therefore disentangling monotop
and anti-monotop production, will be carried out later in this section.

\begin{figure}
\centering
  \vspace{-0.6cm}
  \epsfig{width=0.75\columnwidth,figure=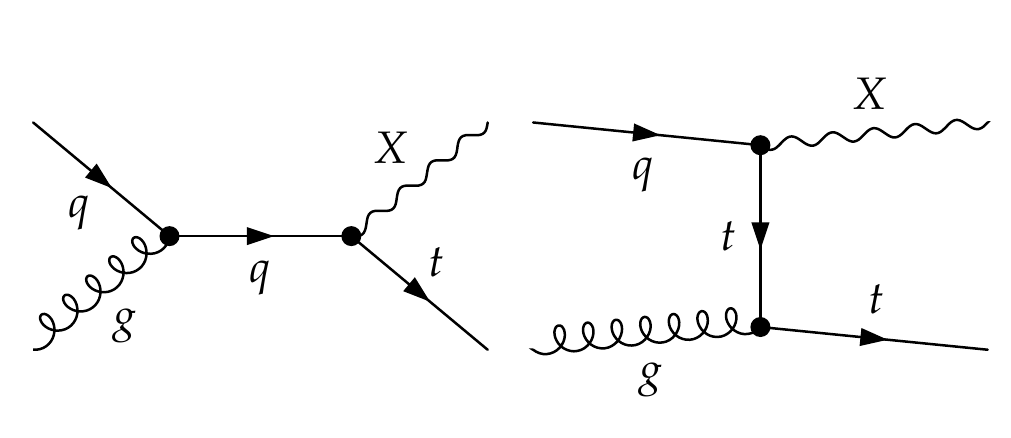}
  \vspace{-0.7cm}
  \caption{\small \label{fig:otherdiag2}Feynman diagrams associated with
 monotop production from
  flavor-changing interactions associated with an extra vector boson $X$ in the $X$--model.}
\end{figure}

The third model for monotop production that we consider
in this work will be denoted by the name `$X$-Model'.
It is motivated by models of dark matter where the top quark
couples to a new neutral vector boson $X_\mu$ strongly interacting with
invisible particles of a hidden
sector~\cite{Kamenik:2011nb, Alvarez:2013jqa}.
The Feynman diagrams for monotop production in this scenario are shown
in Figure~\ref{fig:otherdiag2}.
In this case, the associated production of the new $X$-boson, which
typically
decays into  particles of the hidden sector and thus escapes detection,
with a top quark leads to a monotop signature.
Adopting a simplified approach, we fix the part of the
Lagrangian relevant for monotop production to
\be
  {\cal L}_X = g_X^i \bar u_i \gamma^\mu P_R t X_\mu + {\rm h.c.} \ ,
\label{eq:LX}\ee
where $P_R$ denotes the right-handed chirality projector and $g_X$ the
associated vector of coupling constants in generation space. 
In the following,
we focus on the dominant production channel where the $X$-boson couples
to an up quark and a top quark (see Table~\ref{tab:pdflumis}).
The experimental constraints on the new physics mass scale in this
scenario have been studied in Ref.~\cite{Kamenik:2011nb}.
Their  findings indicate that the mass of the $X$ field can be as low as 100 GeV wihtout
conflicting with current bounds, such as $B_{d,s}-\bar{B}_{d,s}$ mixing
or rare top decays.

The two Lagrangians of Eqs.~\eqref{eq:Lhylo} and \eqref{eq:LX}
allow us to calculate the corresponding
fully polarized partonic cross-sections.
We obtain, using the narrow width approximation for the Hylogenesis case
and providing the differential cross-section with respect to the Mandelstam
$t$-variable for the $X$-model,
\be\label{eq:otherxsec}\bsp
 \hat \sigma^{h_1h_2}_{\rm hylo} (\bar q_j \bar q_k\to t \chi) =&\
   \frac{2 (1-h_1h_2) \pi \big| \kappa_{jk}
     \big|^2}{3} \ \times\ \textrm{BR}\Big(V \to t \chi\Big)\ \times \
   \delta\big(\hat s - m_V^2\big) \ , \\ 
  \frac{{\rm d}\hat \sigma^{h,\lambda}_X}{{\rm d} t} (u_i g \to t X)= &\
    \frac{1}{16 \pi s^2} \frac{g_s^2 g_X^{i2}}{12 s m_X^2 (t-m_t^2)^2}(1+h)
    \Big[ C_1 +  C_2\lambda\Big] \ .
\esp\ee

In the Hylogenesis model,
$h_1,h_2$ are the helicities of the initial
partons and the results for the charge-conjugate processes are obtained by
replacing $h_i \to -h_i$.
It is clear from the fully polarized partonic cross-section of
Eq.~(\ref{eq:otherxsec}) that in this case, the single-spin asymmetries
vanish exactly. In contrast, for the double-polarized asymmetries,
we have a similar situation as for RPV monotop production and the
hadron-level asymmetries can be written in terms of ratios of
the partonic luminosities discussed in Section~\ref{sec:lumipol}.
To be explicit, for the dominant production channel
$\bar{d}\,\bar{d}$+$d\,d$ we have
\be
A_{L}^{\bar{d}\bar{d}+dd} = 0 \, , \qquad
A_{LL}^{\bar{d}\bar{d}+dd} = -\frac{\mathcal{L}^{LL}_{dd}+\mathcal{L}^{LL}_{\bar{d}\bar{d}}}{\mathcal{L}_{dd}+\mathcal{L}_{\bar{d}\bar{d}}} \, ,
\ee
and likewise for other initial states.

In the $X$-model calculation, 
we have kept the dependence on the gluon and initial quark
polarizations $\lambda$ and $h$ explicit and have introduced the kinematical
factors
\be\bsp
 C_1(s,t) =&\ m_t^8 - m_t^6 \Big[2s+t\Big] +m_t^4\Big[ (s+t)^2 - 2 m_X^2(t+m_X^2)\Big]\\
 &\  + m_t^2\Big[ 4m_X^6-2 m_X^4 t + 2 m_X^2(s^2-st+2t^2) - t(s+t)^2\Big]\\
  &\ -  2 m_X^2 t(2m_X^4+s^2+t^2  - 2 m_X^2 (s+t)\Big] \ ,\\
 C_2(s,t) =&\ \Big[m_t^4 \!+\! m_t^2(2m_X^2 \!-\! s \!-\! t) \!+\! 2 m_X^2 (s\!-\!t) \Big]
   \Big[m_t^4 \!+\! t(2m_X^2 \!-\! s \!-\! t) \!-\! m_t^2 (2 m_X^2\!+\!s) \Big] \ .
\esp\ee
The main feature of this class of models with respect to the RPV case lies in
the various helicity combinations contributing to
the polarized cross-section.
Important differences are consequently expected in spin asymmetries 
when comparing RPV monotop production to Hylogenesis or dark matter
$X$-model predictions.

\begin{figure}
\centering
  \epsfig{width=.49\columnwidth,figure=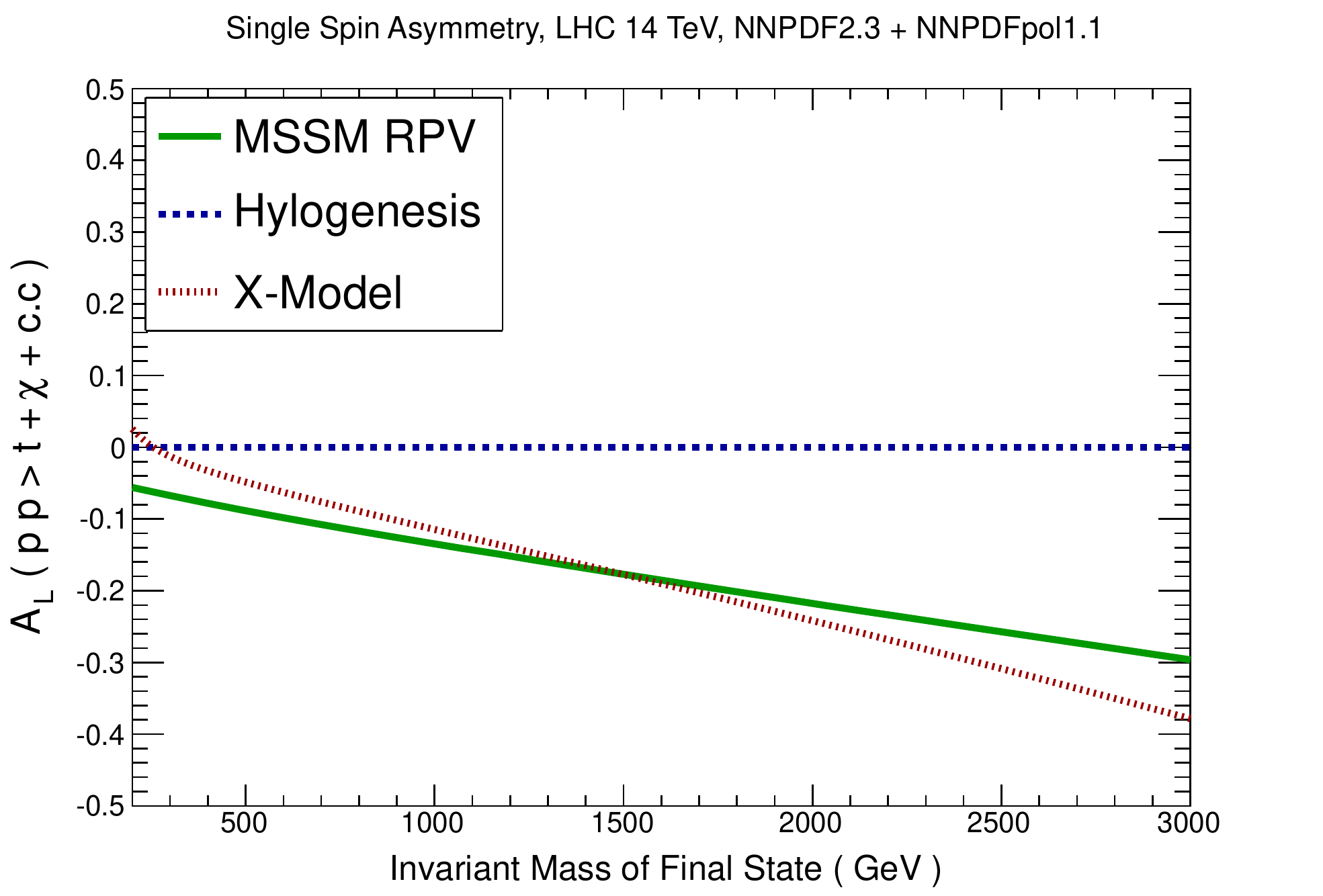}
  \epsfig{width=.49\columnwidth,figure=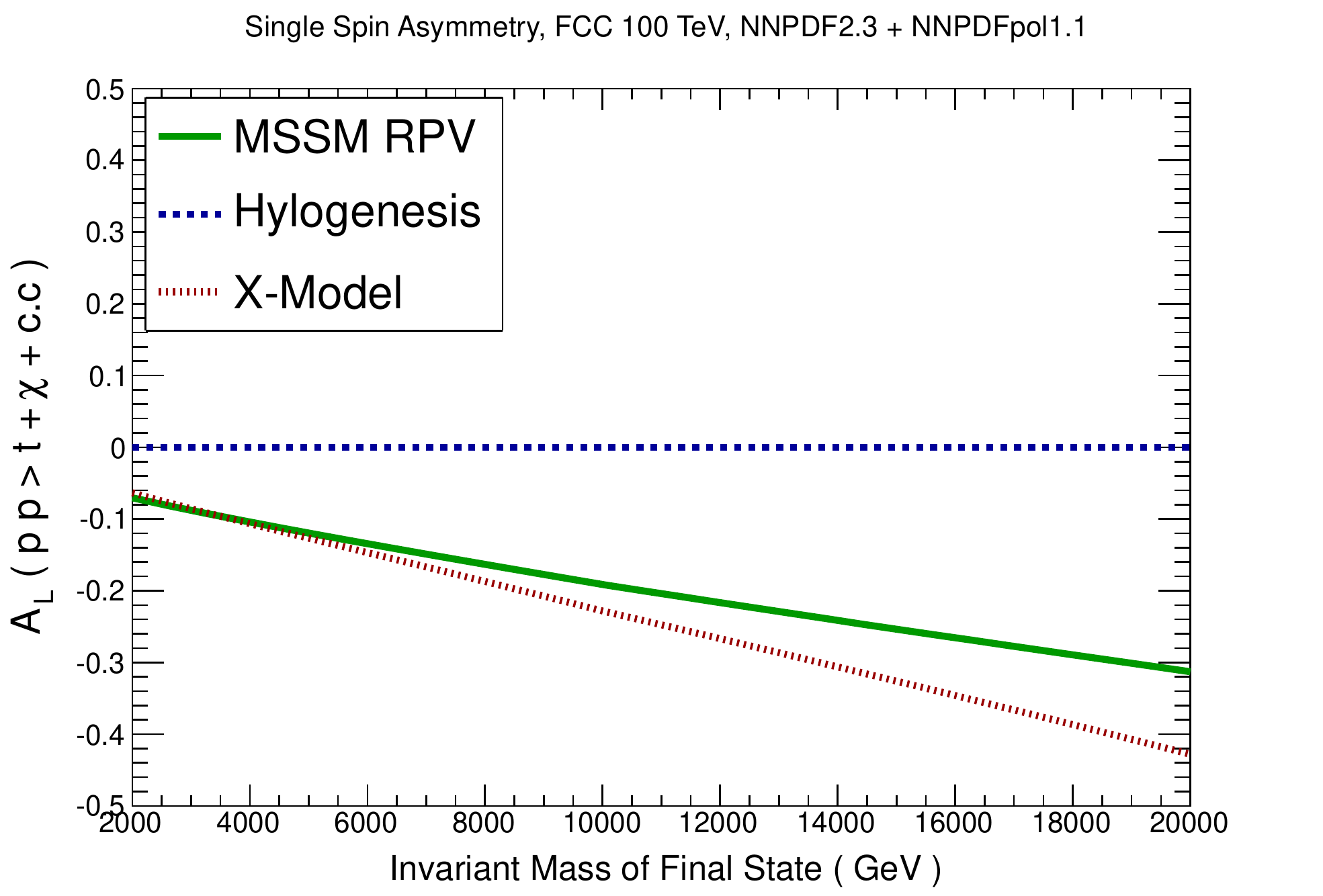}\\\vspace{.3cm}
\epsfig{width=.49\columnwidth,figure=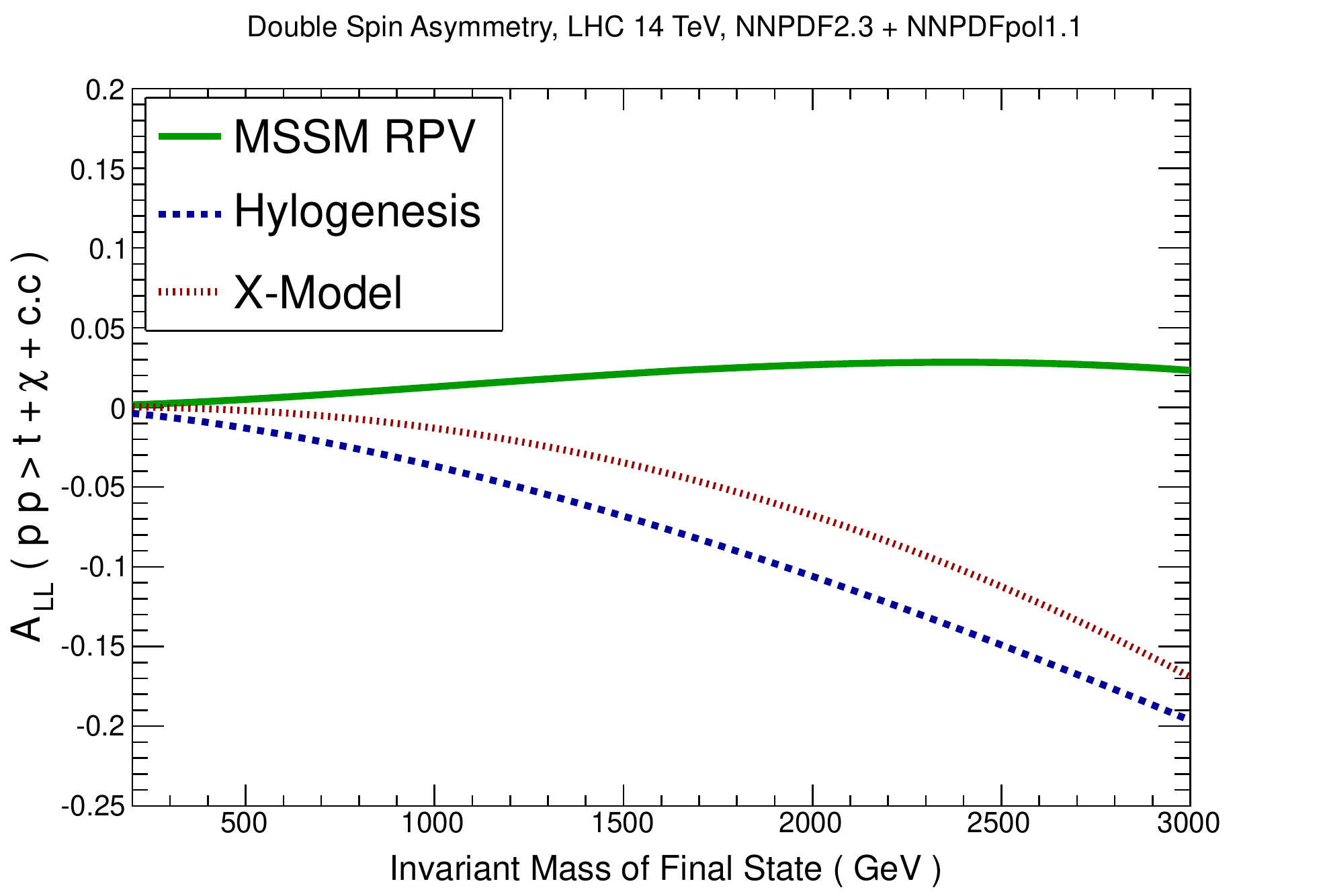}
  \epsfig{width=.49\columnwidth,figure=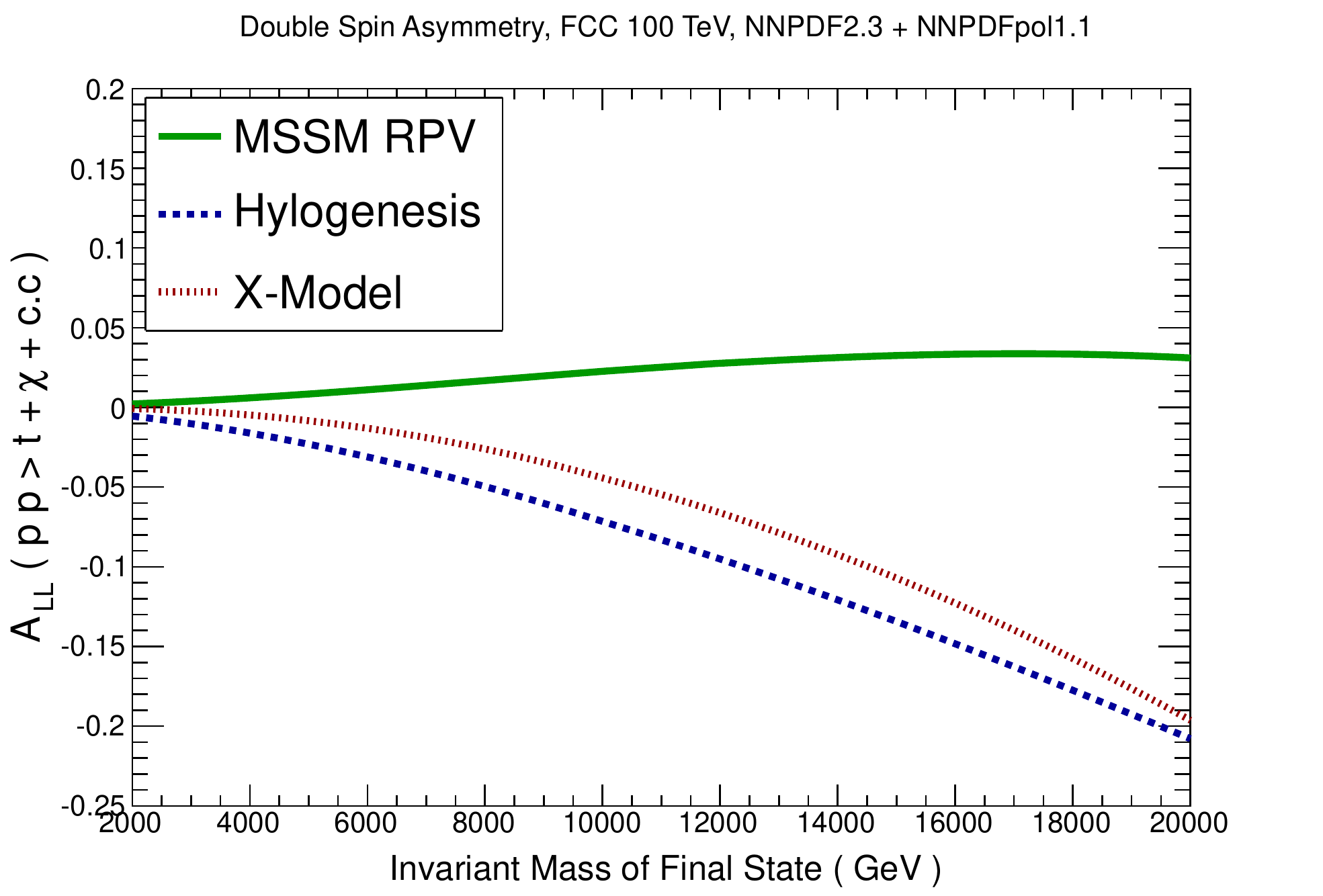}
  \caption{\small \label{fig:otherasym}Single-spin (upper panel) and double-spin (lower panel)
  asymmetries for monotop production at the LHC 14 TeV (left panel) and at an
  FCC 100 TeV (right panel) as function of the monotop system mass for the various
  new physics scenarios described in the text.
  Asymmetries have been obtained using NNPDFpol1.1 and NNPDF2.3.
Sum over monotop and anti-monotop production is implicit.
}
\end{figure}

This is illustrated in Figure~\ref{fig:otherasym} where we compare,
for illustrative purposes,
single-spin and double-spin asymmetries as predicted in RPV scenarios
where the monotop system originates from
a $ds+\bar d\bar s$ initial state, in Hylogenesis models where it
is produced from $dd+\bar d\bar d$ scattering and in dark matter $X$-models
where the $X$-boson arises from the $g\lp u+\bar{u}\rp$ initial state. 
 We present our results as functions of 
the monotop system mass being defined as the resonance mass for both the RPV
and the Hylogenesis scenarios, and as the sum of the $X$-boson and top quark masses
for the $X$-model case\footnote{The numerical values of all other model parameters
are irrelevant as canceling in the ratios of polarized and unpolarized cross-sections.}.

Figure~\ref{fig:otherasym} is the main result of this work.
It tells us that, assuming polarized PDF uncertainties are improved by
a series of dedicated measurements, a measurement
of the single-spin asymmetry with 5\% precision would 
allow one to discriminate between the three production mechanisms
for states with sufficiently large invariant mass, 
approximately above 2 TeV at the LHC and above 10 TeV
at the FCC.
Even a measurement of the sign of the single spin asymmetries would be
very valuable to discriminate between different scenarios.
Double-spin asymmetries would provide a complementary cross-check of the
single-spin results, though their measurement is rather more challenging
both because of the reduced rates and because of the smaller values
of the asymmetries.
The qualitative behavior of $A_L$ and $A_{LL}$ is also found to be
rather different
in some scenarios. In RPV monotop production, for instance, $A_{L}$ is large
and negative, while $A_{LL}$ is small and positive. It is thus clear that a
simultaneous measurement of $A_{L}$ and $A_{LL}$ would provide
stringent constraints on the underlying production dynamics.

\subsection{Impact of monotop charge tagging}

In the final part of this section, we study what we can learn if the charge
of the final-state top quark is tagged, that is, if we are able
to disentangle the monotop
signature (a top quarks of charge +2/3 and missing transverse energy) from the
anti-monotop signature
(same with a top antiquark).
This charge tagging could
be potentially relevant because in these two cases, the polarized PDFs
that are relevant according to the nature of the initial state can show quite
different behaviors.
Tagging the charge of the top quark can thus provide another handle
on the underlying BSM scenario that has induced monotop
production\footnote{Charge asymmetries are also interesting
observables in the context of unpolarized collisions.}.

\begin{figure}
\centering
  \epsfig{width=.49\columnwidth,figure=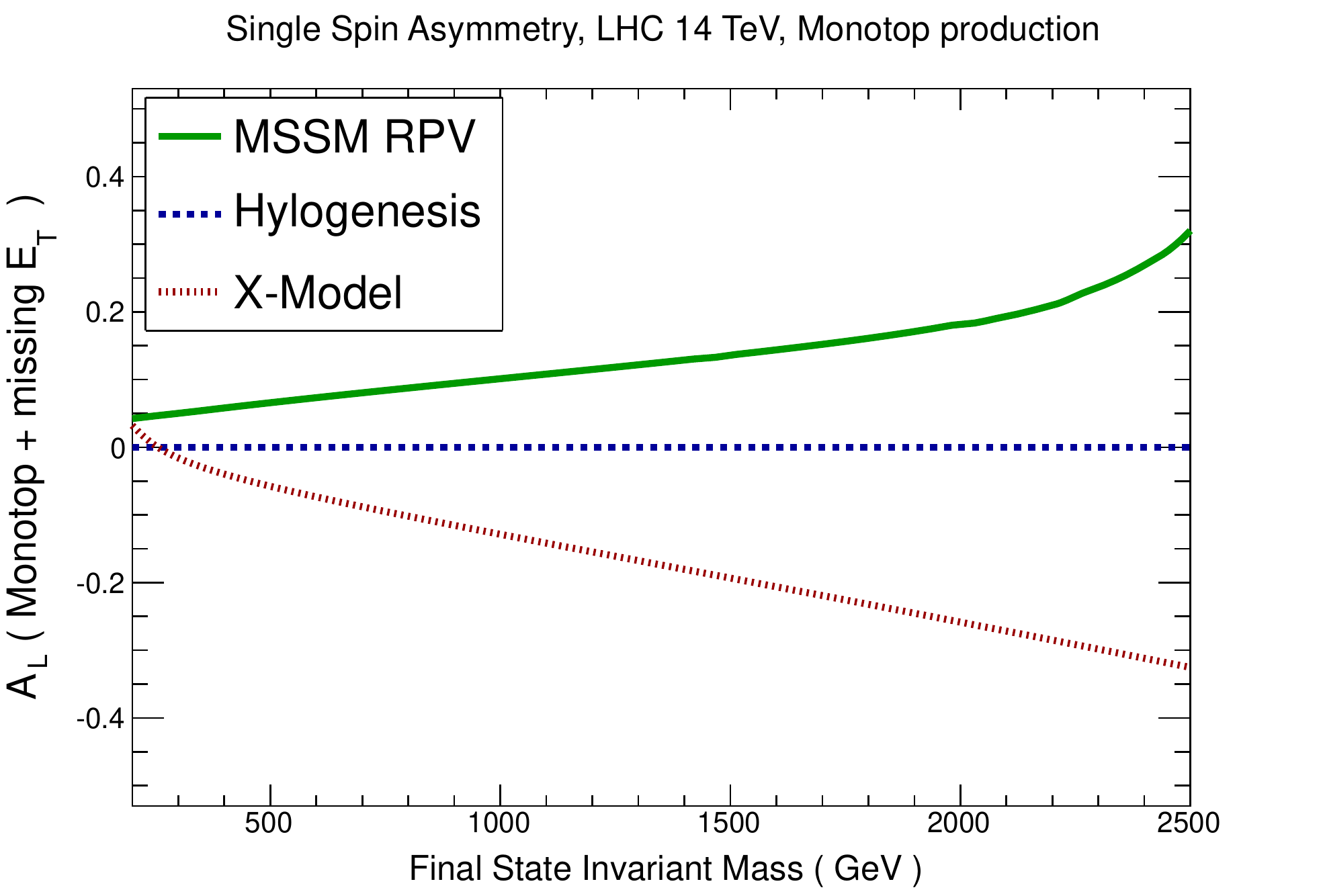}
  \epsfig{width=.49\columnwidth,figure=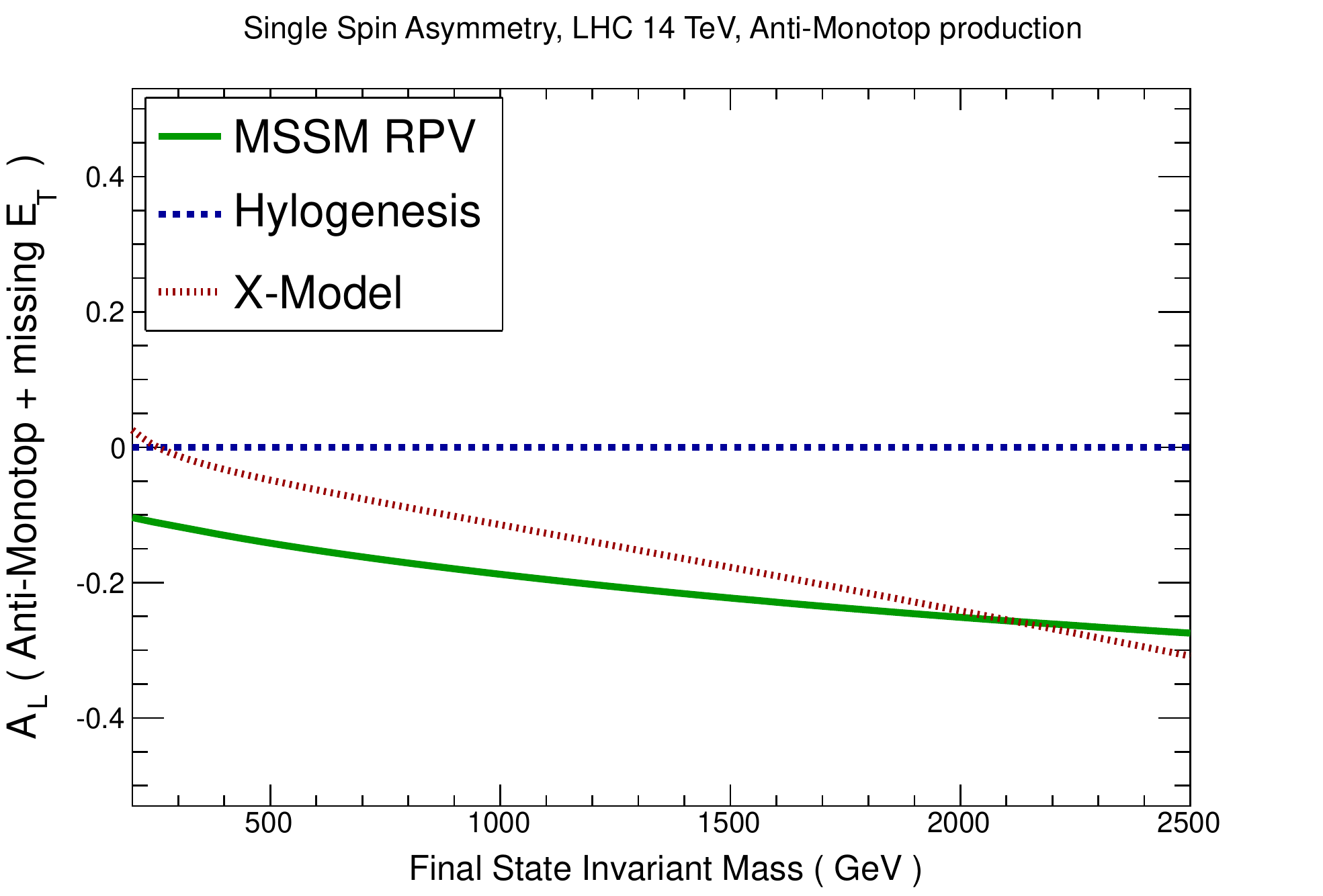}\\\vspace{.3cm}
  \epsfig{width=.49\columnwidth,figure=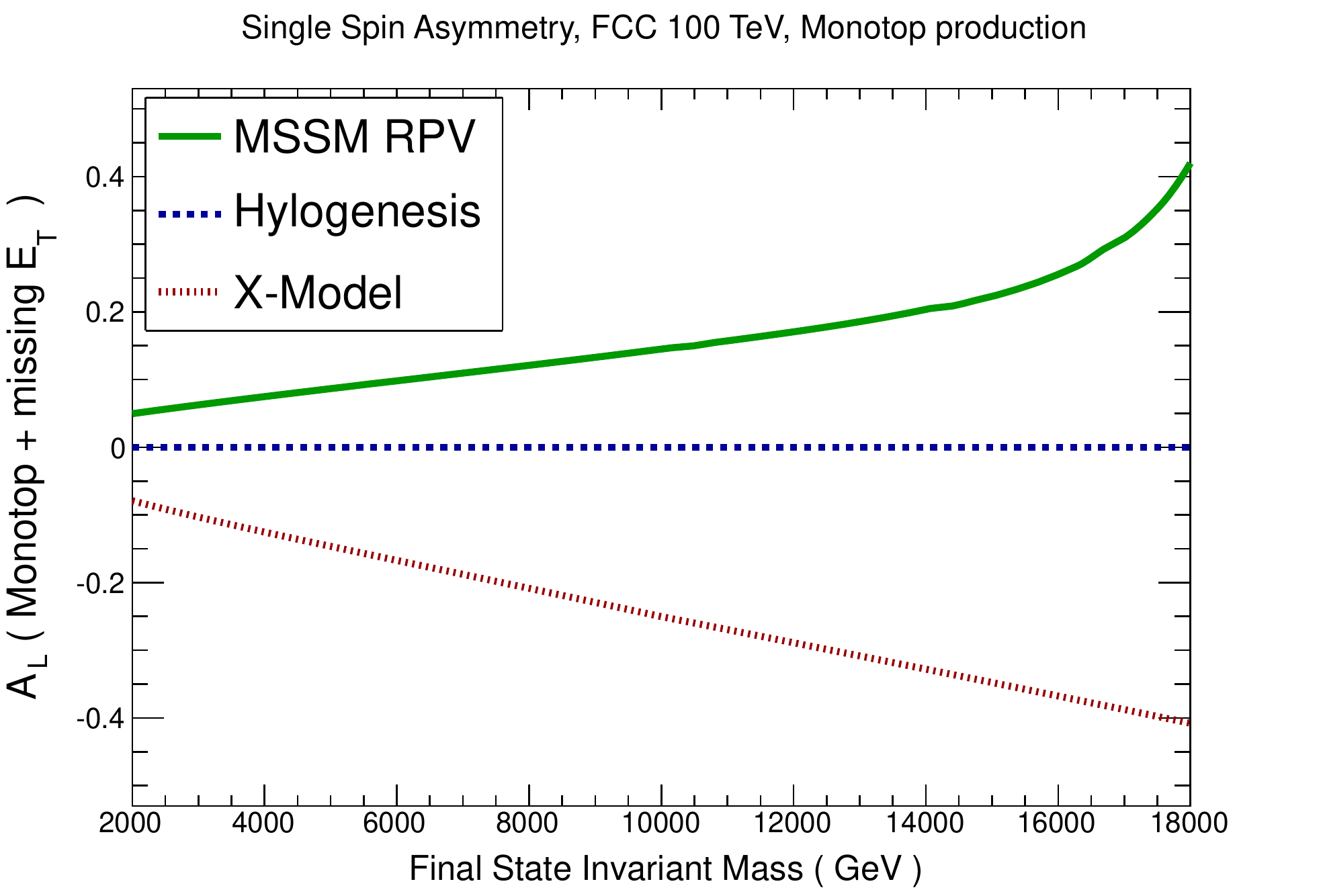}
 \epsfig{width=.49\columnwidth,figure=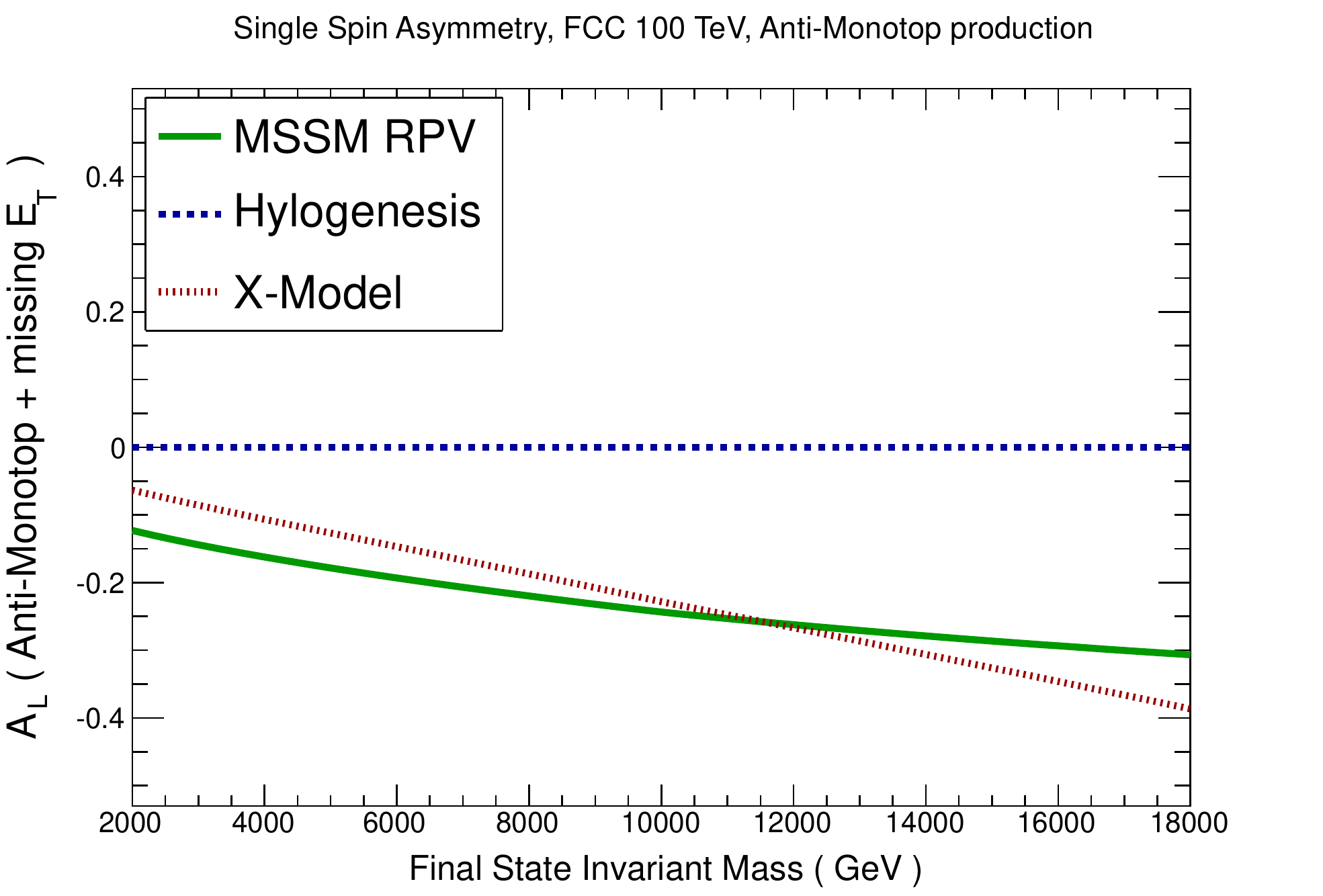}
  \caption{\small \label{fig:otherasymCC} Single-spin 
  asymmetries for monotop (left panel) and anti-monotop (right panel)
production at the LHC 14 TeV (upper row) and at an
  FCC 100 TeV (lower row) as function of the
invariant mass of the final state for the different
  new physics scenarios discussed in this work.
}
\end{figure}

We show in Figure~\ref{fig:otherasymCC} the single-spin asymmetries
for LHC 14 TeV and FCC 100~TeV, this time separating monotop
from anti-monotop production, for the three models under consideration.
The relevant initial states for monotop
production are $\bar{d}\,\bar{s}$,
$\bar{d}\,\bar{d}$ and $u\,g$ in the RPV, Hylogenesis and 
$X$-model scenarios respectively, and the corresponding charge-conjugate
ones for anti-monotop production.
It is clear from the differences between the left and right 
columns of Figure~\ref{fig:otherasymCC} that tagging the top quark
charge provides important information about the underlying production
model, with the differences particularly striking in the case of the
RPV scenario, where at large masses a different sign of the
asymmetry is predicted in the two cases.

We recall that the results of Figure~\ref{fig:otherasym} cannot be
retrieved by a trivial average of the asymmetries of
Figure~\ref{fig:otherasymCC} over monotop and anti-monotop production,
as the full singly-polarized and unpolarized cross sections need to be averaged first,
before evaluating the ratios.

\subsection{Summary}
In this section, we have shown how, using polarized collisions, it is possible
to discriminate between different models that lead to the same
final-state signature, in this case monotop production.
While we have considered this specific benchmark scenario, it is clear
that our results/our considerations 
apply to a wide variety of other BSM models
where the availability of polarized beams would provide a unique
handle for their characterization.

\section{Conclusions and outlook}
\label{sec:conclusions}

In this paper, we have motivated how the availability of polarized beams
at high-energy hadron colliders provides a unique handle on the
discrimination between different beyond the 
Standard Model scenarios that lead to the
same final-state signatures in unpolarized collisions.
First of all, we have discussed in a model-independent way
why single and double-spin asymmetries in polarized collisions
allow us for the separation between different initial-state production mechanisms.
Then we have considered different benchmark scenarios
for monotop production and shown how the measurement of
spin asymmetries in polarized collisions could help to discriminate
between different models.
Therefore, while polarized beams are certainly not required for
BSM discoveries, they can provide very useful information on
the properties of the 
hypothetical BSM sector, in particular in the
determination of its couplings to Standard Model particles.

While technically feasible, the likelihood of a future polarized mode
at the LHC is very small, requiring a complete modification of the full
injector chain.
The situation might however be different 
for the recently proposed Future Circular Collider (FCC) at
a center-of-mass energy of 100~TeV: 
if there is a strong physics case, we believe
that the polarized option should be considered seriously.
In any case this is the right time to begin to think of the feasibility
of such an option, now that various studies for the planning of this
machine have just started.
In particular, if new physics is discovered at the LHC during the proton-proton
runs at center-of-mass energies of 13~TeV, 14~TeV or at the future high-luminosity upgrade
of the LHC, there will be a very
strong motivation for a polarized mode of the FCC
in order to characterize and understand the properties of this new sector.

Future studies, along similar directions as the ones we have explored in this
paper, should be performed in two different and complementary
directions.
On the one hand, other BSM scenarios should be studied.
These studies should focus on the production of high-mass
particles, since as we discussed this is the only region
where single- and double-spin asymmetries are relatively 
large and thus experimentally accessible.
A possible example would be to estimate the accuracy to which the
couplings of a possible heavy $Z'$ can be determined at the FCC
from polarized collisions.
On the other hand, one also needs to perform more detailed feasibility
studies for the measurement of single- and double-spin asymmetries,
trying to estimate the luminosities in the polarized mode that
a 100 TeV FCC could deliver and how the rates would be affected by the finite
polarization of the beams.
Quantifying  the statistical uncertainties of the spin asymmetries at the
FCC would also allow one to better understand what is the reach of BSM characterization
of the polarized collision mode.

As an intriguing final remark, it should be noted that at a 100 TeV FCC it might be possible to access 
polarized collisions without the need of using polarized beams\footnote{We thank G.~Salam
for pointing out this observation to us.}.
Indeed, at the scale of 10-20 TeV, the electroweak $W$- and $Z$-bosons
are effectively massless
and should be included in the DGLAP evolution,
which leads at this point to an intrinsic polarization of the
quarks and gluons via mixing.
This is an interesting possibility to study further, since in any case
PDFs with electroweak corrections are mandatory for the physics of
a 100 TeV hadron collider.

\acknowledgments
The authors are grateful to Emanuele Nocera for useful discussions. This work
has been partially supported by the Th\'eorie-LHC-France initiative of the
CNRS/IN2P3, by the French ANR 12 JS05 002 01 BATS@LHC and by
a PhD grant of the `Investissements d'avenir, Labex ENIGMASS'.

\bibliographystyle{JHEP}
\bibliography{rpvmono}

\end{document}